%% file: paper.tex
\documentclass[12pt]{article}
\pdfoutput=1

\usepackage[pdftex]{graphicx}
\usepackage[nointlimits]{jafhmath}
\usepackage{color}
\usepackage{multirow}
\usepackage{subfigure}
\usepackage{authblk}
\usepackage{amssymb}
\usepackage{latexsym}
\usepackage[centertags]{amsmath}
\usepackage{cite}
\usepackage{fancyhdr}
\usepackage{color}
\usepackage{verbatim}

\input{definitions}

\begin{document}

\title{\bf High-order Discretization of a Gyrokinetic Vlasov Model in
  Edge Plasma Geometry}   
\author[1]{Milo R. Dorr\thanks{Corresponding author, {\tt dorr1@llnl.gov}}}
\author[3]{Phillip Colella}
\author[2]{Mikhail A. Dorf}
\author[1]{Debojyoti Ghosh}
\author[1]{\\Jeffrey A. F. Hittinger}
\author[3]{Peter O. Schwartz}
\affil[1]{\small Center for Applied Scientific Computing, Lawrence Livermore
  National Laboratory, 7000 East Avenue L-561, Livermore, CA 94550
\footnote{This work performed under the auspices of the
  U.S. Department of Energy by Lawrence Livermore National Laboratory
  under Contract DE-AC52-07NA27344.}
}
\affil[2]{Fusion Energy Program, Lawrence Livermore National
  Laboratory, 7000 East Avenue L-630, Livermore, CA 94550}
\affil[3]{Applied Numerical Algorithms Group, Lawrence Berkeley
  National Laboratory, One Cyclotron Road Mail Stop 50A-1148,
  Berkeley, CA 94720
\footnote{Research supported by the Office of Advanced
  Scientific Computing Research of the US Department of Energy under
  contract number DE-AC02-05CH11231.}
}
\date{}

\maketitle

\DeclareGraphicsExtensions{.pdf}

\input{abstract}
\input{introduction}

\input{gksystem}

\input{discretization}
\input{mapping}
\input{cogent}

\input{verification}

\input{summary}

\input{acknowledgments}

\input{appendix}

%\bibliography{journalISI20,publishers,refs}
\bibliography{journalISI20,refs}
\bibliographystyle{plain}

\end{document}

%% file: definitions.tex
\newcommand{\Bpll}{{B_{\parallel}}}
\newcommand{\Bstarpll}{{B^*_{\parallel}}}
\newcommand{\bnabla}{\boldsymbol{\nabla}}
\newcommand{\vpll}{v_{\parallel}}

\newcommand{\mbxi}{\mathbf{\xi}}

%% file: abstract.tex
\begin{abstract}

We present a high-order spatial discretization of a continuum
gyrokinetic Vlasov model in axisymmetric tokamak edge plasma
geometries.  Such models describe the phase space advection of plasma
species distribution functions in the absence of collisions.
The gyrokinetic model is
posed in a four-dimensional phase space, upon which a grid is imposed
when discretized.  To mitigate the computational cost associated with
high-dimensional grids, we employ a high-order discretization to
reduce the grid size needed to achieve a given level of accuracy
relative to lower-order methods.

Strong anisotropy induced by the magnetic field motivates the use of
mapped coordinate grids aligned with magnetic flux surfaces.  The
natural partitioning of the edge geometry by the separatrix between
the closed and open field line regions leads to
the consideration of multiple mapped blocks, in what is known as a
mapped multiblock (MMB) approach.
We describe the specialization of a more general formalism that we have developed for the
construction of high-order, finite-volume discretizations on MMB
grids, yielding the accurate
evaluation of the gyrokinetic Vlasov operator, the metric
factors resulting from the MMB coordinate mappings, and the interaction
of blocks at adjacent boundaries.

Our conservative formulation of the gyrokinetic Vlasov model
incorporates the fact that the phase space velocity has zero
divergence, which must be preserved discretely to avoid truncation
error accumulation.  We describe an approach for the discrete
evaluation of the gyrokinetic phase space velocity that preserves the
divergence-free property to machine precision.

A distinguishing feature of an edge geometry is the X point, where the
poloidal field component vanishes.  The inability to construct fully
flux-surface aligned MMB coordinate systems that are smooth up to and
through the X point requires the relaxation of alignment in the
vicinity of this point.  We therefore describe an approach for the generation of
suitable discrete coordinate block mappings.

The algorithms described in this paper form the foundation of the
COGENT continuum gyrokinetic edge code, which is used here to perform
a convergence study verifying the accuracy of the high-order
spatial discretization.

\end{abstract}

%% file: introduction.tex
\section{Introduction \label{sec:intro}}

\begin{figure}
\centering
\includegraphics[height=4.in]{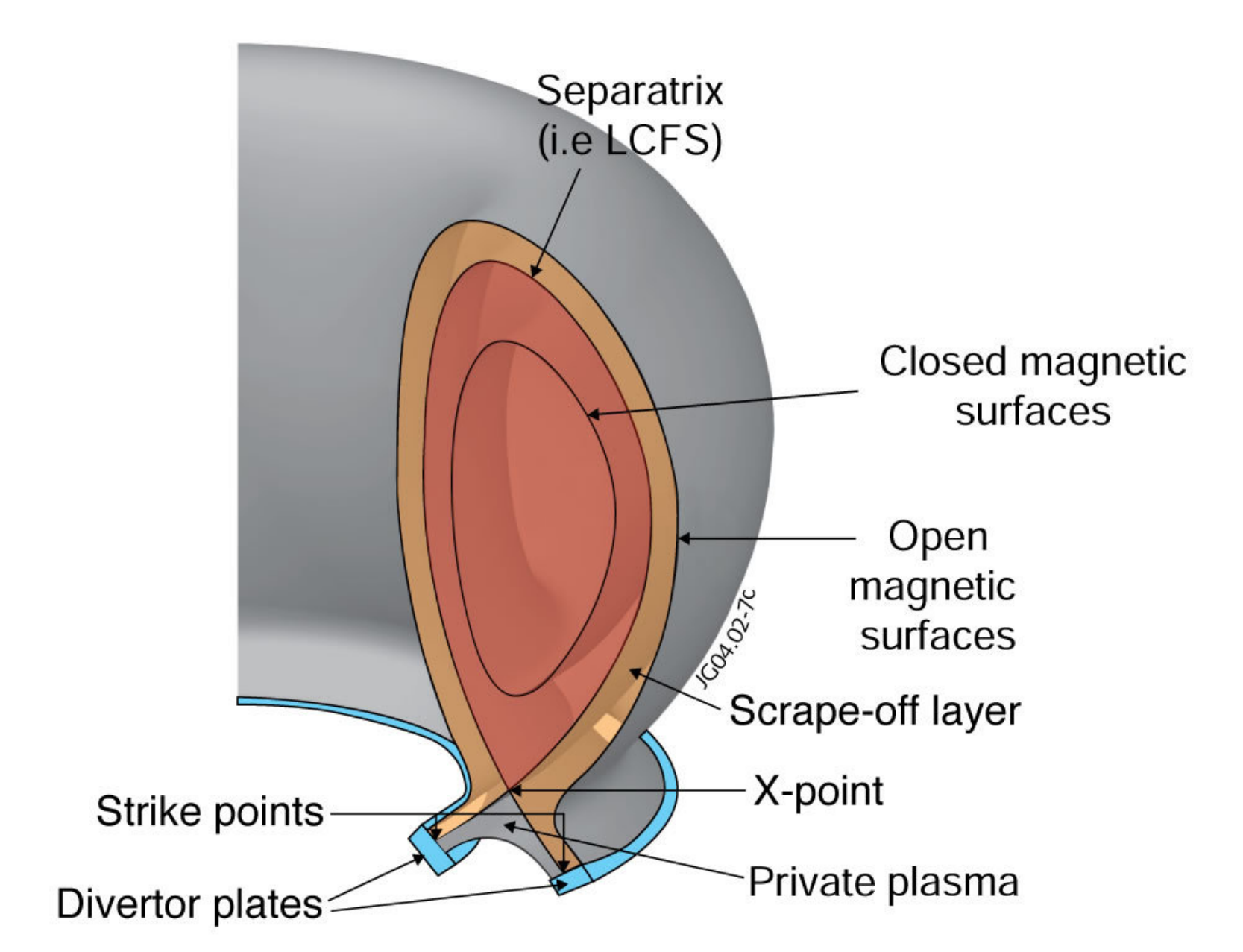} 
\caption{Edge geometry schematic\cite{Euro-fusion}.  The edge region
  is defined from the outer wall, where field lines terminate at the
  divertor plates or at the wall, across the magnetic separatrix and
  into the core region of closed, concentric flux surfaces.}  \label{fig:edge}
\end{figure}

\begin{figure}
\centering
\includegraphics[viewport=0 200 500 400,width=6.5in]{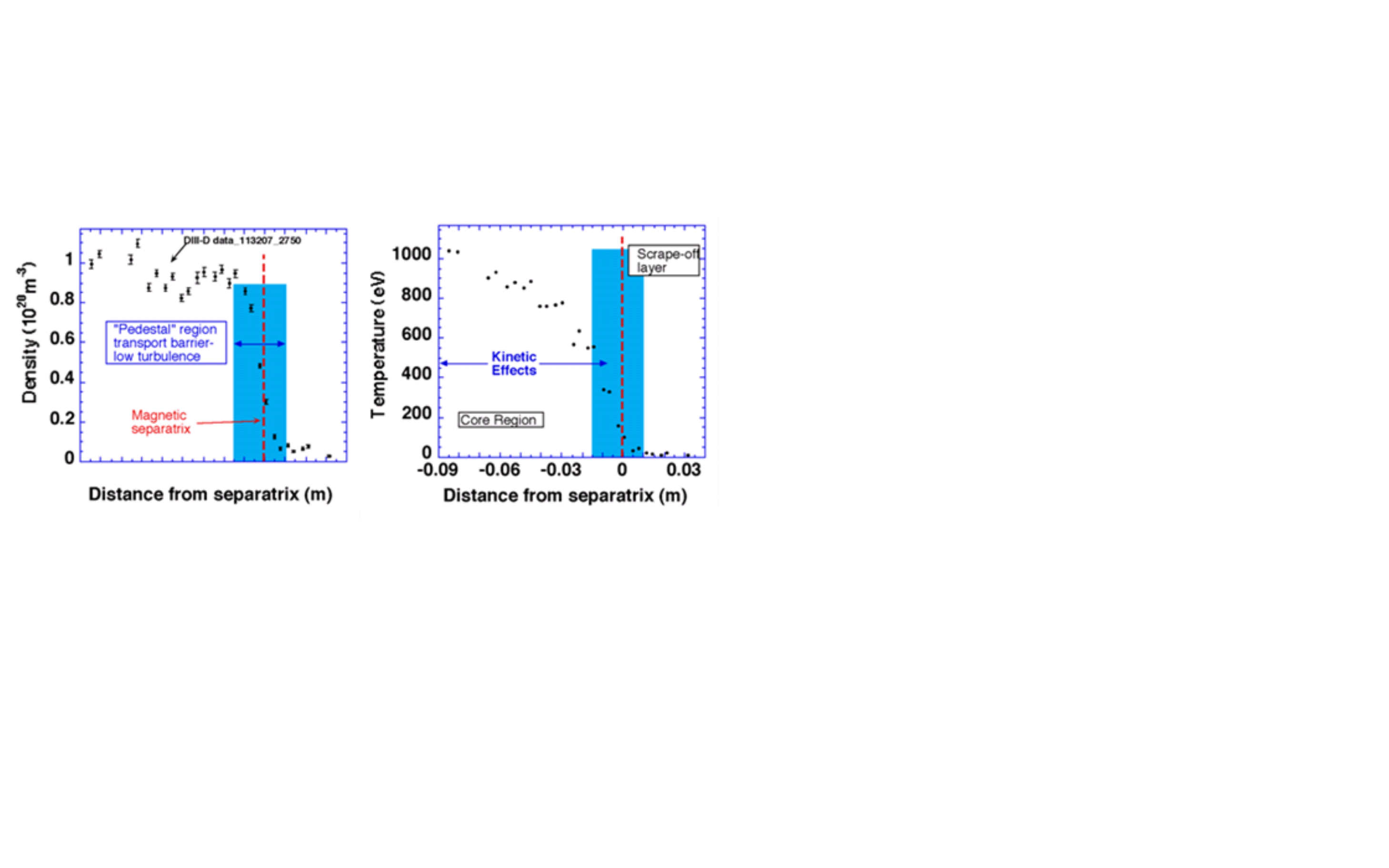} 
\caption{Plasma density and temperature measured in the DIII-D tokamak
  \cite{PoIsBoRo00}.  The region of rapid transition occurring near
  the separatrix is known as the pedestal.}  \label{fig:pedestal}
\end{figure}

The ability to model computationally the behavior of magnetically
confined plasma in the edge region of a tokamak fusion reactor is a
key component in the development of a predictive simulation capability
for the entire device.  As shown schematically in Figure
\ref{fig:edge}, the edge region spans both sides of the magnetic
separatrix, encompassing part of the core region, part or all of the
region between the plasma and the reactor wall, and the divertor
plates.  In addition to geometric complexity, an important feature
that distinguishes the edge from the core is the development of a
region of steep gradients in the density and temperature profiles
called the pedestal (Figure \ref{fig:pedestal}), the height of which
determines the quality of plasma confinement, and hence fusion gain.
A kinetic plasma model is needed in this region, because the radial
width of the pedestal observed in experiments is comparable to the
radial width of individual particle orbits (leading to large
distortions of the local distribution functions from a Maxwellian),
while the mean free path can be comparable to the scale length for
temperature variations along the magnetic field (violating the
assumptions underlying a collisional fluid model).

Because of the large number of independent variables in a fully
kinetic model, as well as the fast time scale represented by the ion
gyrofrequency, gyrokinetic models have been developed that remove the fast gyromotion
asymptotically and thereby facilitate numerical treatments.  Continuum
models consist of a Vlasov operator describing the evolution of plasma
species distribution functions in a particular coordinate system
combined with some variant of Maxwell's equations and collision terms.
Codes such as GENE\cite{Jenko00b,Jenko05},
GS2\cite{Dorland00, Kotschenreuther95b} and
GYRO\cite{GYRO,Candy03a,Candy03b} have been successfully employed to
model core plasmas for many years.  In addition to requiring simpler
geometries, these codes exploit the fact that in the core,
distribution functions are typically small perturbations $\delta \! f$
about a known Maxwellian distribution $f_0$, providing a simpler, and
even sometimes linear, model.  To model the edge plasma all the way to
the reactor walls, a method to solve nonlinear gyrokinetic models for
the entire distribution function (so-called full-$f$) in edge-relevant
geometries is needed.

\begin{figure}
\centering
\setlength{\unitlength}{0.01in}
\begin{picture}(600,350)
\put(-150,-20){\includegraphics[width=5.5in]{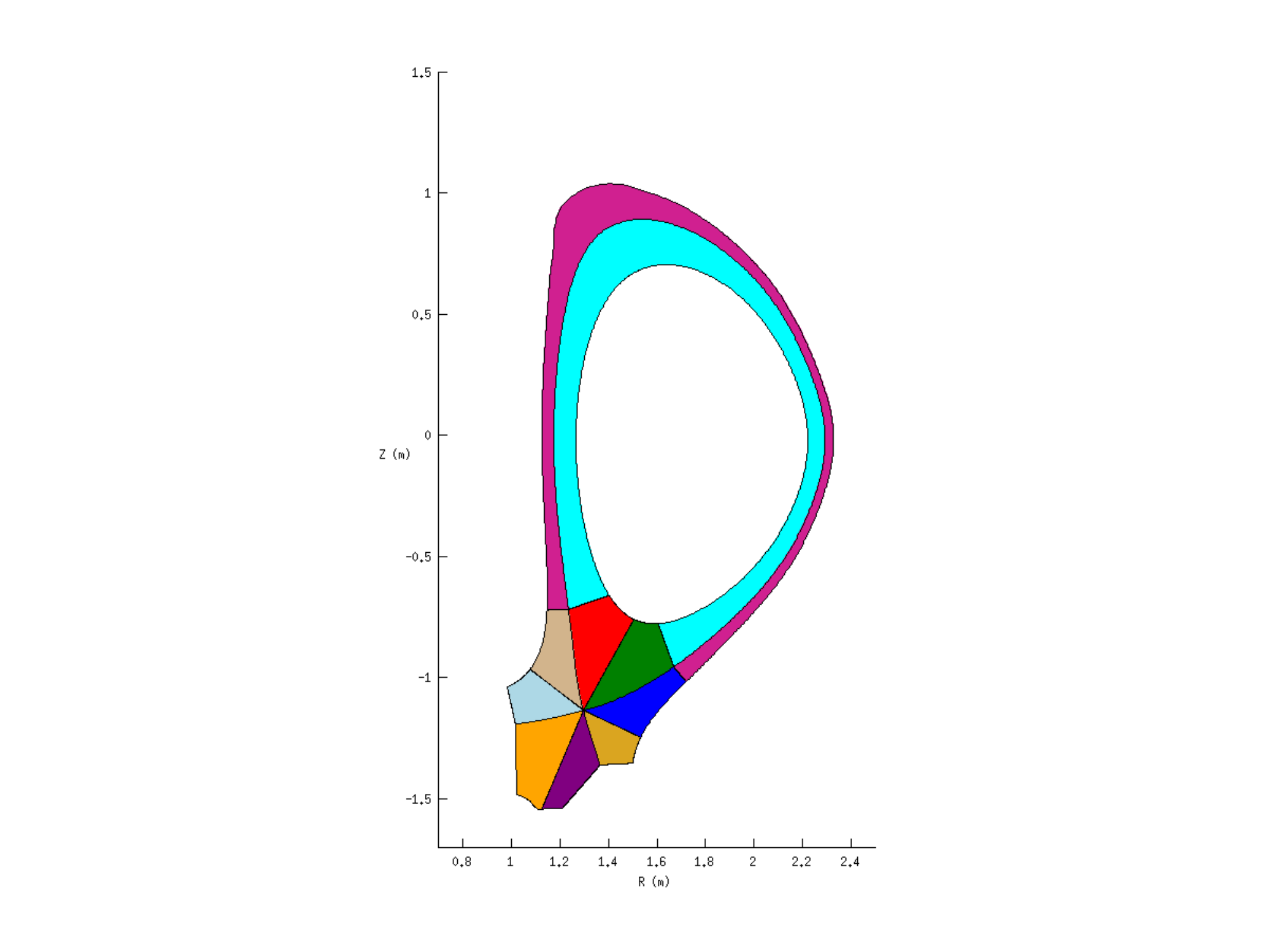}}
\put(200,10){\includegraphics[width=4.5in]{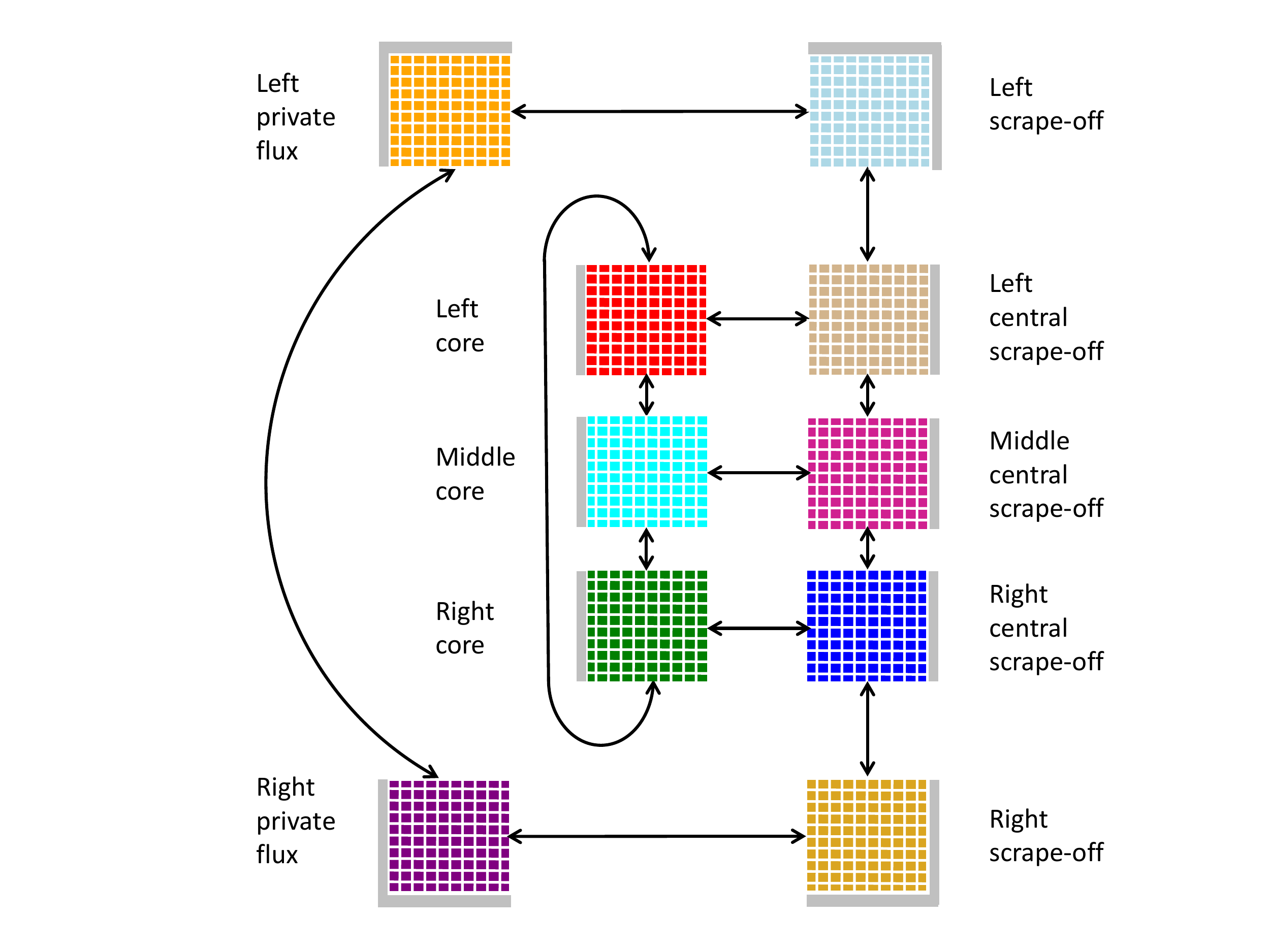}}
\end{picture}
\caption{Mapped multiblock decomposition of the edge geometry (left)
  and the corresponding locally rectangular computational domain (right).  The arrows indicate
the inter-block connectivity, and the gray regions indicate physical boundaries.\label{fig:singlenull}}
\end{figure}

In this paper, we describe an approach for the discretization of a
continuum full-$f$, gyrokinetic Vlasov system in axisymmetric edge
geometries.  The system is treated as a conservation law describing the
advection of plasma species distribution functions in a four
dimensional phase space (2 configuration space and 2 velocity space
coordinates).  A (mostly) flux-surface-aligned, mapped-multiblock
(MMB) coordinate system (Figure \ref{fig:singlenull}) is used to
accommodate strong anisotropy in the
direction of magnetic field lines, similar to fluid edge codes
such as UEDGE \cite{rognlien:1992,Rognlien99,Rognlien02} and SOLPS-ITER
\cite{SOLPS2015}.  We formulate a high-order spatial
discretization to reduce the phase space grid size needed for a
specific level of accuracy relative to a lower-order approximation, as
well as to reduce numerical dissipation in long-time integrations.  In
particular, we employ a semi-discretization based on a general
formalism \cite{CoDoHiMa10,McCoDoHi2013} for the creation of
arbitrarily high-order finite-volume spatial discretizations in mapped
coordinates.  This formalism is summarized in Sections
\ref{sec:singleblock} and \ref{sec:mappedmultiblock}.  A unique
feature of the edge geometry application is that a key requirement of
the formalism, namely, that the block mappings are smooth up to and
some distance beyond block boundaries, cannot be satisfied by a
completely flux-surface-aligned mapping due to the singularity of the
metric factors at the X point, where the magnetic separatrix
intersects itself and the poloidal field component vanishes.  As discussed in Section
\ref{sec:mapping}, some dealignment is therefore necessary in a
neighborhood of the X point to enable high-order differentiation
there.

Since the phase space velocity in gyrokinetic models is developed via
Hamiltonian equations of motion, the zero velocity divergence implied by the
area-preserving property of such dynamical systems is an important
constraint that must be maintained during discretization to avoid the
accumulation of truncation error in long-time integrations.  For the
gyrokinetic model \cite{Ha96} addressed here and presented in
Section \ref{sec:gksystem}, we demonstrate in
Section \ref{sec:divfree} how the divergence-free property can be
satisfied to machine precision in the context of our high-order, MMB,
finite-volume discretization.  Specifically, in calculating the phase
space cell averages of the gyrokinetic flux divergence, the velocity
enters through integrals of the normal components over the cell faces,
which in turn are used to compute flux normals via a high-order
product formula.  By exploiting the fact that the phase space velocity can be
written as a sum of vectors including a certain skew-symmetric,
second-order tensor divergence, Stokes' theorem can be invoked
to reduce the cell face velocity
integrals to face boundary integrals that telescopically sum to zero
in the cell-averaged finite volume divergence calculation.  Additional
benefits accrue from the fact that the face boundary integrals
corresponding to three of the four velocity terms ({\em i.e.},
parallel streaming, curvature and $\nabla B$ drifts) are computed
exactly, and no metric factors appear.

The algorithms described herein provide the foundation of the COGENT
(\underline{CO}ntinuum \underline{G}yrokinetic \underline{E}dge
\underline{N}ew \underline{T}echnology) code, which we have been
developing for the solution of continuum gyrokinetic models in
multiblock geometries, including those describing the tokamak edge.
An overview of COGENT is the topic of Section \ref{sec:cogent}, and in
Section \ref{sec:verification}, we utilize COGENT to demonstrate the
accuracy of the spatial discretization described in the preceding
sections.  The efficient discretization of the gyrokinetic Vlasov
operator is necessary but not sufficient for the complete simulation
of the edge plasma problem.  Given its fundamental importance,
however, we have limited the scope of this paper to only that part.
As described in Section \ref{sec:cogent}, COGENT includes a number of
the other components ({\em e.g.}, collision operators, self-consistent
electrostatic fields and high-order semi-explicit time integration
methods) needed to address edge-relevant problems, which have been
used in several verification studies
\cite{DoCoCoDoHi10,DoEtAl2012,DorfEtAl2013,CoEtAl2013,DorfEtAl2016,DorfEtAl2014,Ghosh2017}.
To our knowledge, COGENT's ability to solve a
continuum gyrokinetic model in edge geometries spanning both sides of
the magnetic separatrix is unique.

%% file: gksystem.tex
\section{The gyrokinetic Vlasov system \label{sec:gksystem}}

We target a reduced version of the full-$f$ gyrokinetic model of \cite{Ha96}:
\begin{equation}
  \frac{\partial (\Bstarpll f)}{\partial t} + 
  \bnabla_{\mbR} \cdot  \left( \dot{\mbR} \Bstarpll f \right) 
  + \frac{\partial}{\partial v_{\parallel}} \left(
    \dot{\vpll} \Bstarpll f \right) = 0,
  \label{conservativenormalized}
\end{equation}
where
\begin{subequations}
\label{gkvelocity}
\begin{eqnarray}
  \dot{\mbR} &=&
  \dot{\mbR}(\mbR,v_{\parallel},\mu,t)
  =
  \frac{v_{\parallel}}{\Bstarpll}\mbB^*+\frac{\rho_L}{Z\Bstarpll}
  \mbb \times \mbG,
  \label{dotterms1normalized} \\
\dot{\vpll} &=& \dot{\vpll}(\mbR,v_{\parallel},\mu,t) =
-\frac{1}{
  m \Bstarpll}
  \mbB^* \cdot\mbG, \label{dotterms2normalized}
\end{eqnarray}
\end{subequations}
and
\begin{subequations}
\label{gkvelocity_vars}
\begin{eqnarray}
  \mbB^*  &= &   \mbB^*(\mbR,v_{\parallel}) = \mbB +
  {\rho_L}\frac{m v_{\parallel}}{Z}
  \bnabla_{\mbR} \times  \mbb, \\
\Bstarpll & = & \Bstarpll (\mbR,v_{\parallel})  = \mbb \cdot  \mbB^*, \\
  \mbG & = &    \mbG
  (\mbR,\mu, t) = Z \bnabla_{\mbR} \Phi + \frac{\mu}{2} \bnabla_{\mbR}
  B.   \label{Gdef}
\end{eqnarray}
\end{subequations}
The unknown quantity $f = f(\mbR,\vpll,\mu,t)$ is
the plasma species distribution function in
gyrocenter phase space coordinates $(\mbR,\vpll,\mu)$, which are described
further below and whose equations
of motion are given by (\ref{gkvelocity})-(\ref{gkvelocity_vars}). 
$B$ and $\mbb$ are the magnitude and direction of the magnetic field $\mbB
= B\mbb$, respectively.  $Z$
and $m$ are the species charge state and mass, respectively.  $\Phi$ is the electric
potential. For our present purposes, we assume the long wavelength limit in which
the Larmor radius is much smaller than the characteristic length
scales for electrostatic potential variations.
We utilize a particular
normalization described in Appendix \ref{normalizations} that nondimensionalizes all quantities,
including the Larmor number, $\rho_L$, defined in Table \ref{table:dimlessnums}.

Gyrocenter coordinates play a key role in gyrokinetic models in two
important ways.  First, they reduce what would otherwise be a
six-dimensional phase space to five dimensions: $\mbR$ is the
three-dimensional configuration space coordinate, $\vpll$ is the
velocity space component along field lines, and the magnetic moment
$\mu = m v_\perp^2 / 2B$ is related to the velocity
$v_\perp$ perpendicular to field lines.  Through the use of asymptotic
orderings, gyrocenter coordinates are specifically constructed so as
to make the distribution function $f$ symmetric with respect to
gyrophase.  The latter component, which would have been the third
velocity component, can then be ignored.  The magnetic moment $\mu$,
an adiabatic invariant, is assumed to be constant in the development
of gyrokinetic theories, which is why no evolution equation appears
for it.  The second benefit of gyrocenter coordinates is that the
gyrofrequency is eliminated, which would otherwise represent a fast
time scale that would need to be resolved.  

The gyrokinetic phase space velocity
(\ref{gkvelocity}) models
strong flow along magnetic field lines, represented by the $\vpll
\mathbf{B}$ quantity, together with curvature, $\nabla B$ and
$\mathbf{E} \times \mathbf{B}$ drift terms containing the
$\rho_L$ factor, which is assumed small in the gyrokinetic
asymptotic ordering.  For typical tokamak parameters, the difference in the
magnitude of the streaming and drift terms can be a few orders of
magnitude.  The resulting strong anisotropy resulting from the
disparity in parallel and perpendicular quantities has many
implications in the development of numerical algorithms.  Because gyrocenter
coordinates are developed as a Hamiltonian dynamical system, the velocity
(\ref{gkvelocity}) satisfies the area preserving property
\begin{equation}
  \bnabla_{\mbR} \cdot \left( \Bstarpll \dot{\mbR} \right) 
  + \frac{\partial}{\partial {v}_{\parallel}} \left(
    \Bstarpll \dot\vpll \right) = 0,  \label{veldiv}
\end{equation}
where $\Bstarpll$ is the Jacobian of the
mapping between lab frame and gyrocenter coordinates.
The analytic verification of
(\ref{veldiv}) is contained in Appendix \ref{sec:veldiv}.  As noted in \cite{Ha96}, the
gyrokinetic Vlasov equation can therefore be expressed in either
convective or conservative form.  We choose the latter with the
objective of achieving a correspondingly conservative numerical
discretization.  The potential $\Phi$ in (\ref{Gdef}) is evaluated by solving some form
of Maxwell's equations.  For the purposes of this paper, we assume
that $\Phi$ is known, although the COGENT code used for our numerical
example includes additional options, as discussed in Section
\ref{sec:cogent}.

The gyrokinetic system is posed in a domain defined by the tokamak
magnetic geometry, which is comprised of field lines lying on
concentric flux surfaces (Figure \ref{fig:edge}).  In this paper, we assume an axisymmetric
geometry, where all quantities are assumed to be constant in the
toroidal direction.  The configuration space domain therefore consists
of a single poloidal slice.  Due to large variations of plasma
parameters along and across field lines (and therefore along and
across sliced flux surfaces in the poloidal plane), there is strong
motivation to discretize in coordinates where at least one of the
coordinate directions is defined by the flux surfaces.  Since a
single, smooth, flux-surface-aligned coordinate mapping cannot be
constructed over the entire domain, we consider a multiblock
decomposition such as the one depicted in the left-hand side of Figure
\ref{fig:singlenull}.  This decomposition is constructed by first
recognizing the natural partitioning defined by the magnetic
separatrix into the core, scrape-off and private flux regions.  Each
region is further decomposed into blocks such that ({\em i\/}) each
block can be mapped from a logically rectangular computational domain
and ({\em ii\/}) adjacent blocks must abut along entire boundaries.
The core region is therefore subdivided into the left (LCORE), middle
(MCORE) and right (RCORE) blocks; the scrape-off layer is decomposed
into the left (LSOL), left-central (LCSOL), middle-central (MCSOL),
right-central (RCSOL) and right (RSOL) blocks, and the private flux
region is decomposed into the left (LPF) and right (RPF) blocks.  The
right-hand side of Figure \ref{fig:singlenull} depicts the inter-block
connectivity in the mapped coordinate domain.  Within each mapped
block, rectangular grids are introduced, resulting in a block
rectilinear gridding of the physical domain.  We require that the
grids be conformal across inter-block boundaries, but otherwise no
additional grid smoothness across block boundaries is assumed.  For
parallelization purposes, the rectangular block grids are further
decomposed in 4D, load balanced and assigned to processors in a
fully general manner.

We comment that the multiblock decomposition rules imposed above could
have been satisfied using only six blocks by combining the LCORE,
MCORE and RCORE blocks into a single block and the LCSOL, MCSOL and
RCSOL blocks into another block.  However, the generation of smooth
mappings on the resulting ``long and skinny'' merged blocks is
non-trivial.  The advantage of the ten-block
decomposition shown in Figure \ref{fig:singlenull} is that eight of
the blocks are curvilinear quadrilaterals that are relatively modest
deformations of the unit square; the MCORE and MCSOL blocks can be
mapped to the unit square using nearly polar coordinate
transformations.

On this mapped multiblock grid, we consider the discretization of the
system (\ref{conservativenormalized})-(\ref{gkvelocity_vars}).  Among our requirements
is the discrete enforcement of the zero velocity divergence condition
(\ref{veldiv}) assumed in the conservative formulation.  A second
requirement is high-order ({\em i.e.}, greater than second-order)
accuracy, which reduces the number of phase space degrees of freedom
to achieve a given level of accuracy.  A high-order method is
also important for reducing numerical dissipation in long-time
integrations.  We therefore begin with a review of a general
approach \cite{CoDoHiMa10} for constructing fourth-order, finite-volume
discretizations of hyperbolic conservation laws in mapped coordinates
on a single block.  In this context, we then describe the calculation
of the gyrokinetic velocities in the mapped coordinate system such
that the divergence free condition (\ref{veldiv}) is satisfied to
machine precision.  In our finite volume approach, the mapped
grid normal velocities are used in flux calculations on cell faces
together with discretized distribution functions.  On cell faces near
interblock boundaries, calculation of the latter is accomplished using suitably
high-order interpolation \cite{McCoDoHi2013}, enabling the extension of the discretization
to multiblock geometries such as the edge geometry in Figure
\ref{fig:singlenull}.

%% file: discretization.tex
\section{Phase space discretization \label{sec:discretization}}

Fundamental to our approach is a general strategy for the systematic development of
high-order, finite-volume discretizations in mapped coordinates.  Additional
details are contained in \cite{CoDoHiMa10}.

\subsection{Fourth-order, finite-volume discretization in a mapped block \label{sec:singleblock}}

Consider a smooth mapping $\mbX$ from the unit cube
onto the spatial domain $\Omega$:
\begin{gather*} 
  \mbX = \mbX(\mbxi) \hbox{,  }
  \mbX:[0,1]^D \rightarrow \Omega.
\end{gather*}
Given this mapping, the divergence of a vector field on $\Omega$
can be written in terms of derivatives in $[0,1]^D$, which
will serve as our computational domain. That is,
\begin{subequations}
\begin{gather}
  \nabla_{\mbX} \cdot \mbF = J^{-1} \nabla_{\mbxi} \cdot
  (\mbN^T \mbF),
  \label{eqn:divmapped} \\
    J = \det\left(\frac{\partial{\mbX}}{\partial\mbxi}\right), ~~ (\mbN^T)_{p, q} = \det\left(R_p
  \left(\frac{\partial{\mbX}}{\partial\mbxi}, \mbe^q\right)\right), ~
  1 \le p,q \le D, \label{metrics}
\end{gather}
\end{subequations}
where $R_p (\mbM, \mbv)$ denotes the matrix obtained by
replacing the $p^{\rm th}$ row of the matrix $\mbM$ by the vector $\mbv$,
and $\mbe^d$ denotes the unit vector in the $d^{\rm th}$ coordinate
direction.

In a finite volume approach, $\Omega$ is
discretized as a union of control volumes.   For Cartesian grid finite volume methods, a
control volume $W_{\mbi}$ takes the form 
\begin{gather*}
  W_{\mbi} = \left [ \mbi h ,
  \left ( \mbi + 
  \mbu \right )h \right ] 
  \hbox{ , } \mbi \in \mathbb{Z}^D \hbox{ , }
  \mbu = (1,1,\ldots,1),
\end{gather*}
where $h$ is the grid spacing.  When using mapped coordinates, we
define control volumes in $\Omega$ as the images
$\mbX(W_{\mbi})$ of the
cubic control volumes $W_{\mbi} \subset [0,1]^D$.
Then, by changing variables and applying the divergence theorem, we obtain the flux
divergence integral over a physical control volume $\mbX(W_{\mbi})$ by
\begin{align}
  \int \limits_{{\mbX(W_{\mbi})}} \nabla_{\mbx} \cdot \mbF d \mbx =  
  \int \limits_{W_{\mbi}} \nabla_{\mbxi} \cdot (\mbN^T
  \mbF) d \mbxi 
  =  \sum \limits^{D-1}_{d = 0} \sum_{\alpha = 0}^1 (-1)^{\alpha+1} \int \limits_{V^\alpha_d} 
  (\mbN^T \mbF)_d dV_{\mbxi},
  \label{eqn:mappedDivergence}
 \end{align}
where the $V_d^0$ and $V_d^1$ are lower and upper faces of cell $W_{\mbi}$ in
the $d$-th direction, respectively.  As described in \cite{CoDoHiMa10}, the integrals on
the cell faces $V^\alpha_d$ can be
approximated using the following formula for the average of a product
in terms of fourth-order accurate face averages of each factor:
\begin{equation}
\left < ab \right >_{\mbi+\frac{1}{2}\mbe^d} = \left < a \right
>_{\mbi+\frac{1}{2}\mbe^d} \left < b \right >_{\mbi+\frac{1}{2}\mbe^d} +
\frac{h^2}{12} \mbG_0^{\perp,d} \left ( \left < a \right
>_{\mbi+\frac{1}{2}\mbe^d} \right ) \cdot \mbG_0^{\perp,d} \left (
\left < b \right >_{\mbi+\frac{1}{2}\mbe^d} \right ) + O(h^4). \label{productaverage}
\end{equation}
Here, $\mbG^{\perp,d}_0$ is the second-order accurate central difference
approximation to the component of the gradient operator orthogonal to
the $d$-th direction: $\mbG^{\perp,d}_0 \approx \nabla_{\mbxi} -
\mbe^d \frac{\partial}{\partial \xi_d}$, and
the operator
$\langle \cdot \rangle_{\mbi + \frac{1}{2}
\mbe^d}$ denotes a fourth-order accurate average over the
cell face centered at $\mbi + \frac{1}{2}
\mbe^d$:
\begin{gather*}
\langle q \rangle_{\mbi + \frac{1}{2} \mbe^d} = \frac{1}{h^{D-1}} \int \limits_{V_d^\alpha} q d V_{\mbxi} + O(h^4).
\end{gather*}
Alternative expressions to (\ref{productaverage}) are obtained
by replacing the averages $\left < a \right
>_{\mbi+\frac{1}{2}\mbe^d}$ and/or $\left < b \right
>_{\mbi+\frac{1}{2}\mbe^d}$ used in the transverse gradients
$\mbG^{\perp,d}_0$ by the corresponding face-centered pointwise values
$a_{\mbi+\frac{1}{2}\mbe^d}$ and/or $b_{\mbi+\frac{1}{2}\mbe^d}$, respectively.

When applied in the discretization of the phase space divergence
operator in the gyrokinetic system (\ref{conservativenormalized}), we
consider the flux $\mbF = f\mbu$, where
\begin{equation}
\mbu = (u_0, u_1, u_2, u_3) = (u_{\vpll},\mbu_\mbR)
= (\Bpll^* \dot{\vpll},\Bpll^* \dot{\mbR}), \label{velocity}
\end{equation}
and the species
subscript $\alpha$ is dropped for brevity.  We therefore obtain
\begin{equation}
\int \limits_{\mbX(W_{\mbi})} \nabla_{\mbx} \cdot \mbF d\mbx = h^3 \sum_{d=0}^3 \sum_{\alpha=0}^1
(-1)^{\alpha+1} F_{\mbi\pm \frac{1}{2}\mbe^d}^d + O(h^4),  \label{averagedivergence}
\end{equation}
where, taking $(\mbN^T \mbu)_d$ and $f$ as the factors in the product formula (\ref{productaverage}), 
\begin{equation}
F_{\mbi\pm \frac{1}{2}\mbe^d}^d = \left < (\mbN^T \mbu)_d \right
>_{\mbi+\frac{1}{2}\mbe^d}
\left < f \right >_{\mbi+\frac{1}{2}\mbe^d} + \frac{h^2}{12}
\left ( \mbG_0^{\perp,d} \left < (\mbN^T \mbu)_d \right
>_{\mbi+\frac{1}{2}\mbe^d}) \right ) \cdot \left ( \mbG_0^{\perp,d}(
\left < f \right >_{\mbi+\frac{1}{2}\mbe^d}) \right ). \label{fddef_divfree}
\end{equation}

We note that an alternative discretization can be obtained by preserving
the flux $\mbF$ as one of the product formula factors, yielding
\begin{equation}
F_{\mbi\pm \frac{1}{2}\mbe^d}^d = \sum_{s=0}^3 \left < N_d^s \right
>_{\mbi+\frac{1}{2}\mbe^d}
\left < F^s \right >_{\mbi+\frac{1}{2}\mbe^d} + \frac{h^2}{12} \sum_{s=0}^3
\left ( \mbG_0^{\perp,d} \left < N^s_d \right
>_{\mbi+\frac{1}{2}\mbe^d}) \right ) \cdot \left ( \mbG_0^{\perp,d}(
\left < F^s \right >_{\mbi+\frac{1}{2}\mbe^d}) \right ), \label{fddef_fs}
\end{equation}
where $F^s$ is the $s$-th component of $\mbF$ and $N_d^s$ is
the $(s,d)$-th element of the matrix $\mbN$.  In
\cite{CoDoHiMa10}, it is demonstrated that the computation of the face averages $ \left < N_d^s \right
>_{\mbi+\frac{1}{2}\mbe^d}$ can be reduced to integrals over
cell edges.  Moreover, assuming that the edge integrals are performed
with the same quadratures wherever they appear, 
\begin{equation}
  \sum \limits^3_{d=0} \sum \limits_{\alpha=0}^1 (-1)^{\alpha+1} \int
  \limits_{V_d^\alpha} {{N}}^s_d dV_{\mbxi} = 0, \label{cancel}
\end{equation}
which guarantees the freestream property that the divergence of a constant
vector field computed by (\ref{eqn:mappedDivergence}) is identically
zero.  Free-stream preservation is an extremely important property in the
simulation of flows using mapped coordinate systems, since
it represents a constraint on the approximation of the metric terms that
reduces the dependence of computed solutions on the choice of mapping and
metric discretization \cite{Ko06}.
However, as will be demonstrated in the next section, the factorization (\ref{fddef_divfree})
enables a discretization in which the divergence of the gyrokinetic
velocity $\mbu$ vanishes to machine precision, thereby enforcing the
assumption (\ref{veldiv}) made in the conservative formulation
(\ref{conservativenormalized}).  Moreover, the formulation is free of
discretized metric terms, which naturally achieves one of the goals
addressed in free-stream preserving approaches.

Calculation of the face-averaged fluxes (\ref{fddef_divfree}) is
therefore reduced to the calculation of face-averaged distribution
functions and mapped normal velocity components.  One choice for the
former is obtained from the fourth-order, centered-difference formula
\begin{equation}
  \langle{f}\rangle_{\mbi+\frac{1}{2}\mbe^d} =
  \frac{7}{12}\left(\bar{f}_{\mbi}+\bar{f}_{\mbi+\mbe^d}\right) - 
    \frac{1}{12}\left(\bar{f}_{\mbi+2\mbe^d}+\bar{f}_{\mbi-\mbe^d}\right)
    + O(h^4),
    \label{fface}
\end{equation}
where $\bar{f}_{\mbi}$ denotes the average of $f$ on cell $\mbi$.
This results in a dissipationless scheme to which a limiter can also
be added.  Alternatively, an upwind method with order higher than four,
such as the fifth-order WENO scheme \cite{weno5}, may be employed.  Boundary conditions are also implemented
here through the setting of inflow conditions in physical
boundary ghost cells.   Fourth-order face averages of the
mapped normal velocity components can be computed directly from
(\ref{gkvelocity}) using the
product formula (\ref{productaverage}) and metric factor face
averages, but as shown in the next section, a more careful exploitation of the
specific structure of the gyrokinetic velocity results in a
discretization that is also discretely divergence free.

\subsection{Gyrokinetic velocity discretization \label{sec:divfree}}

As shown in Appendix \ref{sec:veldiv}, the divergence of the gyrokinetic velocity
(\ref{gkvelocity}) is zero.  
In this section, we demonstrate how to preserve this property when
computing the divergence using the mapped-grid,
finite-volume discretization described in the preceding section.  Specifically, we
show how the mapped normal velocity components $\left < (\mbN^T \mbu)_d \right
>_{\mbi+\frac{1}{2}\mbe^d}$ can be computed such that the divergence integral
(\ref{eqn:mappedDivergence}) with $\mbF = \mbu$ is zero to
machine precision.

Let $\mbA$ denote a magnetic potential ({\em i.e.}, $\mbB =
\nabla \times \mbA$) and let
\begin{equation}
p =  \rho_L \left ( \Phi + \frac{\mu B}{2Z} \right ). \label{pdef}
\end{equation}
Letting $(x_0,x_1,x_2,x_3) = (\vpll,\mbR)$, we define
\begin{align}
\widetilde{\mbu} = \mbu - \widehat{\mbu},
\end{align}
where $\widehat{\mbu} = (\widehat{\mbu}_j)$ is the vector with entries
\begin{align}
\widehat{\mbu}_j = - \delta_{0,j} \frac{Z}{m\rho_L} \mbB \cdot
\nabla_{\mbR} p, ~~~~ 0 \le j \le 3.
\end{align}
$\widehat{\mbu}$ is divergence-free, and therefore so is $\widetilde{\mbu}$.  
Using the components of $\widetilde{\mbu} =
(\widetilde{u}_0,\widetilde{u}_1,\widetilde{u}_2,\widetilde{u}_3)$, we
define the 3-form
\begin{align}
\bsigma &= \widetilde{u}_3 ~dx_0 \wedge dx_1 \wedge dx_2 - \widetilde{u}_2 ~ dx_0 \wedge dx_1 \wedge dx_3 \\
\nonumber  &+ \widetilde{u}_1 ~ dx_0 \wedge dx_2 \wedge dx_3 - \widetilde{u}_0 ~ dx_1 \wedge dx_2 \wedge dx_3.
\end{align}
The mapped normal velocity components of $\widetilde{\mbu}$ can be
expressed (see Appendix \ref{section:face_integrals})
in terms of surface integrals of $\bsigma$ using the parameterization
given by our mapping $\mbX$ (which implies a particular surface
orientation that has already been accounted for in
(\ref{eqn:mappedDivergence})) as
\begin{align}
  \int \limits_{V^\alpha_d} (\mbN^T \widetilde{\mbu})_d d\mbV_{\bxi} =
  (-1)^{d+1} \int \limits_{\mbX(V^\alpha_d)} \bsigma, \hspace{0.3 in} 0 \le d \le 3. \label{normalvelint}
\end{align}
Since $\widetilde{\mbu}$ is divergence-free, its exterior derivative
$d\bsigma$ is zero.  Poincar\'{e}'s Lemma (applied on the contractible
manifolds $\mbX(V^\alpha_d)$) therefore guarantees the existence of a 2-form
\begin{align}
\bomega = \sum \limits_{0 \le i < i' \le 3} \omega_{i,i'} ~dx_i \wedge dx_{i'}
\end{align}
such that $\bsigma = d\bomega$.  It therefore follows that
\begin{align}
\widetilde{\mbu}_j = \sum \limits_{j' = 0}^3 \frac{\partial
    \zeta_{j,j'}}{\partial x_{j'}}, ~~~~ 0 \le j \le 3, \label{divtensor}
\end{align}
where
\begin{align}
\bzeta = (\zeta_{i,j}) = \left (
\begin{array}{cccc}
0 & - \omega_{0,1} & \omega_{0,2} & - \omega_{0,3} \\
\omega_{0,1} & 0 & - \omega_{1,2} & \omega_{1,3} \\
- \omega_{0,2} & \omega_{1,2} & 0 & - \omega_{2,3} \\
\omega_{0,3} & - \omega_{1,3} & \omega_{2,3} & 0
\end{array}
\right ). \label{zetadef}
\end{align}
The form $\bomega$ is not unique.  As described in Appendix
\ref{sec:veldecomp}, particular $\omega_{i,j}$ satisfying
(\ref{divtensor})-(\ref{zetadef}) can be found by a careful inspection of
$\widetilde{\mbu}$, yielding
\begin{subequations}
\begin{align}
\omega_{0,j} &= (-1)^{j+1}\vpll (\mbb \times \nabla_{\mbR} p )_j, ~~ 1
\le j \le 3, \\
\omega_{1,2} &= - \vpll \left ( \mbA + \frac{
  m \vpll\rho_L}{Z} \mbb \right )_3 , \label{omega12}\\
\omega_{1,3} &= -\vpll \left ( \mbA + \frac{
  m \vpll\rho_L}{Z} \mbb \right )_2 , \\
\omega_{2,3} &= -\vpll \left ( \mbA + \frac{
  m \vpll\rho_L}{Z} \mbb \right )_1 , \label{omega23}\\
\omega_{i,j} &= - \omega_{j,i}, \hspace{1.3in} 0 \le j < i \le 3.
\end{align}
\end{subequations}

Next, letting $\mbX^*$ denote the pullback mapping \cite{diff_forms} defined by $\mbX$, we
apply Stokes' theorem \cite{spivak:calculusOnManifolds} to obtain
\begin{align}
\int \limits_{\mbX(V^\alpha_d)} \bsigma = \int \limits_{V^\alpha_d}
\mbX^*(d\bomega) = \int \limits_{V^\alpha_d} d(\mbX^*\bomega) = \sum
\limits_{d' \ne d} \sum \limits_{\beta =0,1} (-1)^{d'+1+\beta} \int \limits_{A^{\alpha,\beta}_{d,d'}} \mbX^*\bomega, \label{omegaint}
\end{align}
where $A^{\alpha,0}_{d,d'}$ and $A^{\alpha,1}_{d,d'}$ are the low and high side faces of $V^\alpha_d$ in direction $d'$, respectively.
The pullback form is defined by
\begin{align}
  \mbX^*\bomega = \sum \limits_{0 \le i < i' \le 3} ~ \sum \limits_{0 \le j < j' \le 3} \omega_{i,i'}
  \det
  \begin{array}{|cc|}
    \frac{\partial X_i}{\partial \xi_j} & \frac{\partial X_i}{\partial \xi_{j'}} \\
    \frac{\partial X_{i'}}{\partial \xi_j} & \frac{\partial X_{i'}}{\partial \xi_{j'}}
  \end{array}
~  d\mbxi_j \wedge d\mbxi_{j'}.
\end{align}
In terms of coordinates, the pullback integrals are then \cite{diff_forms}
\begin{align}
  \int \limits_{A^{\alpha,\beta}_{d,d'}} \mbX^*\bomega = 
  \sum \limits_{\stackrel{\scriptstyle 0 \le j < j' \le 3}{j,j' \ne d,d'}}
~\sum \limits_{0 \le i < i' \le 3}
\int \limits_{A^{\alpha,\beta}_{d,d'}} 
\omega_{i,i'}
  \det
  \begin{array}{|cc|}
    \frac{\partial X_i}{\partial \xi_j} & \frac{\partial X_i}{\partial \xi_{j'}} \\
    \frac{\partial X_{i'}}{\partial \xi_j} & \frac{\partial X_{i'}}{\partial \xi_{j'}}
  \end{array}
  ~  d\mbxi_j d\mbxi_{j'}. \label{pullbackints}
\end{align}

For example, consider a cylindrical coordinate mapping of the form:
\begin{subequations}
\begin{align}
  X_0 &= v_{\parallel}(\xi_0), \\
  X_1 &= R(\xi_1,\xi_2) \cos(\xi_3), \\
  X_2 &= R(\xi_1,\xi_2) \sin(\xi_3), \\ 
  X_3 &= Z(\xi_1,\xi_2),
\end{align}
\end{subequations}
where $\xi_n^0 \le \xi_n \le \xi_n^1$ for $0 \le n \le 2$ and $0 \le
\xi_3 \le 2\pi$.  The evaluation of (\ref{pullbackints}) yields
\begin{equation}
\int \limits_{A^{\alpha,\beta}_{d,d'}} \mbX^*\bomega =
Q_{d,d'}^{\alpha,\beta},
\end{equation}
where,
\begin{subequations}
\begin{align}
Q^{\alpha,\beta}_{0,1} &= 2 \pi \rho_L \int_{\xi_2^0}^{\xi_2^{1}}
\left \{ \left [ \left ( -\eta_0 (\mbb \times \mathbf{E})_Z + \frac{\eta_1}{2Z}
 (\mbb \times \nabla B)_Z \right ) \frac{\partial R}{\partial \xi_2} \right. \right.\\
&\hspace{6.3em} \left. \left.
 + \left ( \eta_0 (\mbb \times \mathbf{E})_R - \frac{\eta_1}{2Z} (\mbb \times \nabla B)_R \right ) \frac{\partial Z}{\partial \xi_2} 
\right ] R \right \}_{\xi_1 = \xi_1^\beta} d\mbxi_2, \\
Q^{\alpha,\beta}_{0,2} &= 2 \pi \rho_L \int_{\xi_1^0}^{\xi_1^{1}}
\left \{ \left [ \left ( -\eta_0 (\mbb \times \mathbf{E})_Z + \frac{\eta_1}{2Z}
 (\mbb \times \nabla B)_Z \right ) \frac{\partial R}{\partial \xi_1} \right. \right.\\
&\hspace{6.3em} \left. \left.
 + \left ( \eta_0 (\mbb \times \mathbf{E})_R - \frac{\eta_1}{2Z} (\mbb \times \nabla B)_R \right ) \frac{\partial Z}{\partial \xi_1} 
\right ] R \right \}_{\xi_2 = \xi_2^\beta} d\mbxi_1, \\
Q^{\alpha,\beta}_{0,3} &= 
\rho_L \int_{\xi_1^0}^{\xi_1^1}  \int_{\xi_2^0}^{\xi_2^1}
\left [ - \eta_0 \left ( \mbb \times \mbE \right )_\Phi +
  \frac{\mu \eta_1}{2Z} (\mbb \times \nabla B)_\Phi \right ] \left ( \frac{\partial
  R}{\partial\xi_2}\frac{\partial Z}{\partial \xi_1}
 - \frac{\partial R}{\partial \xi_1}\frac{\partial Z}{\partial \xi_2} \right )
 d\mbxi_1 d\mbxi_2, \\
Q^{\alpha,\beta}_{1,2} &= - 2 \pi \left ( \eta_2 R \mbA_\Phi + \eta_3\frac{m \rho_L}{Z}
R\/ b_\Phi \right )_{\xi_1=\xi_1^\alpha,
  \xi_2=\xi_2^\beta} \label{Q12phys}, \\
Q^{\alpha,\beta}_{1,3} &= -\int_{\xi_2^0}^{\xi_2^1}
\left [ \left ( \eta_2 \mbA_R + \eta_3 \frac{m\vpll \rho_L}{Z} \mbb_R \right ) \frac{\partial
  R}{\partial \xi_2} + \left ( \eta_2 \mbA_Z + \eta_3 \frac{m\vpll \rho_L}{Z} \mbb_Z  \right ) \frac{\partial Z}{\partial \xi_2}
\right ]_{\xi_1 = \xi_1^\alpha} d\mbxi_2,
\end{align}
\begin{align}
Q^{\alpha,\beta}_{2,3} &= -\int_{\xi_2^0}^{\xi_2^1}
\left [ \left ( \eta_2 \mbA_R + \eta_3 \frac{m\vpll \rho_L}{Z} \mbb_R \right ) \frac{\partial
  R}{\partial \xi_1} + \left ( \eta_2 \mbA_Z + \eta_3 \frac{m\vpll \rho_L}{Z} \mbb_Z  \right ) \frac{\partial Z}{\partial \xi_1}
\right ]_{\xi_2 = \xi_2^\alpha} d\mbxi_1, \\
Q^{\alpha,\beta}_{d',d} &= -Q^{\alpha,\beta}_{d,d'}
\end{align}
\end{subequations}
and $\eta_0 = \vpll(\xi_0) $, $\eta_1 = \vpll(\xi_0)
\mu$, $\eta_2 = (\vpll^2(\xi_0^1) - \vpll^2(\xi_0^0))/2$
and $\eta_3 = (\vpll^3(\xi_0^1) - \vpll^3(\xi_0^0))/3$.
Noting that the quantities $Q^{\alpha,\beta}_{d,3}$ are independent of $\beta$
for all $d$, we have from (\ref{omegaint}) that
\begin{subequations}
\begin{align}
\int \limits_{\mbX(V^\alpha_0)} \bsigma &= \sum \limits_{d' \ne 0}
\sum \limits_{\beta = 0,1} (-1)^{d' + 1+\beta} 
Q_{0,d'}^{\alpha,\beta} = \sum \limits_{\beta = 0,1} (-1)^\beta
\left ( Q_{0,1}^{\alpha,\beta} - Q_{0,2}^{\alpha,\beta} \right ), \\
\int \limits_{\mbX(V^\alpha_1)} \bsigma &= \sum \limits_{d' \ne 1}
\sum \limits_{\beta = 0,1} (-1)^{d' +1+ \beta} 
Q_{1,d'}^{\alpha,\beta} = \sum \limits_{\beta = 0,1} (-1)^\beta
\left ( Q_{0,1}^{\alpha,\beta} - Q_{1,2}^{\alpha,\beta} \right ), \\
\int \limits_{\mbX(V^\alpha_2)} \bsigma &= \sum \limits_{d' \ne 2} \sum \limits_{\beta = 0,1} (-1)^{d'+1+\beta} 
Q_{2,d'}^{\alpha,\beta} = \sum \limits_{\beta = 0,1} (-1)^\beta
\left ( Q_{0,2}^{\alpha,\beta} - Q_{1,2}^{\alpha,\beta} \right ), \\
\int \limits_{\mbX(V^\alpha_3)} \bsigma &= \sum \limits_{d' \ne 3}  \sum \limits_{\beta = 0,1} (-1)^{d'+1+\beta} 
Q_{3,d'}^{\alpha,\beta} = 0,
\end{align}
\end{subequations}
which, together with (\ref{normalvelint}), yields
\begin{align}
\int \limits_{{\mbX(W)}} \nabla_{\mbX} \cdot \widetilde{\mbu}~ d\mbx =  
\sum \limits^3_{d = 0} \sum \limits_{\alpha = 0,1}  (-1)^{1 + \alpha +
d} \int \limits_{\mbX(V^\alpha_d)} \bsigma  = 0.  \label{zeroutildedivergence}
\end{align}
Trivially,
\begin{align}
\int \limits_{{\mbX(W)}} \nabla_{\mbX} \cdot \widehat{\mbu}~ d\mbx =  
\sum \limits_{\alpha = 0,1} (-1)^{1+\alpha} \int \limits_{V^\alpha_0} 
N_{0,0} \widehat{\mbu}_0 d\mbV_{\bxi} = 0,
\end{align}
and therefore $\mbu$ is discretely divergence free.

The assumption of axisymmetric fields allows additional
simplifications of the quantities $Q_{0,1}$, $Q_{0,2}$, and $Q_{1,2}$.
Assuming that $\mathbf{E}$ has no toroidal component, we observe that
\begin{align}
-(\mbb \times \mathbf{E})_Z \frac{\partial R}{\partial \xi_2}  + (\mbb
\times \mathbf{E})_R \frac{\partial Z}{\partial \xi_2} = -b_\Phi \left
( \frac{\partial \phi}{\partial R} \frac{\partial R}{\partial \xi_2} +
\frac{\partial \phi}{\partial Z} \frac{\partial Z}{\partial \xi_2}
\right ) = -\frac{B_\Phi}{B} \frac{\partial \phi}{\partial \xi_2},
\end{align}
and similarly, since $\nabla B$ has no toroidal component,
\begin{align}
(\mbb \times \nabla B)_Z \frac{\partial R}{\partial \xi_2} - (\mbb
\times \nabla B)_R \frac{\partial Z}{\partial \xi_2} = -
\frac{B_\Phi}{B} \frac{\partial B}{\partial \xi_2} = - B_\Phi
\frac{\partial \ln(B)}{\partial \xi_2}.
\end{align}
Hence, since $RB_\Phi = (RB)_{\rm tor}$ is constant,
\begin{align}
Q^{\alpha,\beta}_{0,1} &= - 2 \pi \rho_L (RB)_{\rm tor} \left [
\eta_0 \int_{\xi_2^0}^{\xi_2^{1}} \left ( \frac{1}{B} \frac{\partial
  \phi}{\partial \xi_2} \right )_{\xi_1=\xi_1^\beta} d\xi_2
 + \frac{\eta_1}{2Z}
\ln \frac{B(\xi_1^\beta,\xi_2^1)}{B(\xi_1^\beta,\xi_2^0))}
\right ]. \label{Q01}
\end{align}
Similarly,
\begin{align}
Q^{\alpha,\beta}_{0,2} &= - 2 \pi \rho_L (RB)_{\rm tor} \left [
\eta_0 \int_{\xi_1^0}^{\xi_1^{1}} \left ( \frac{1}{B} \frac{\partial
  \phi}{\partial \xi_1} \right )_{\xi_2=\xi_2^\beta} d\xi_1
 + \frac{\eta_1}{2Z}
\ln \frac{B(\xi_1^1,\xi_2^\beta)}{B(\xi_1^0,\xi_2^\beta)}
\right ]. \label{Q02}
\end{align}
Letting $\Psi$ denote the magnetic flux, we may take $\mbA_\Phi =
\Psi/R$, yielding
\begin{align}
Q^{\alpha,\beta}_{1,2} &= - 2 \pi \left ( \eta_2 \Psi + \eta_3\frac{m \rho_L}{Z}
\frac{(RB)_{\rm tor}}{B} \right )_{\xi_1=\xi_1^\alpha, \xi_2=\xi_2^\beta}. \label{Q12}
\end{align}
Summarizing the above, the face integrals of the mapped normal
velocities are
\begin{subequations}
\begin{align}
 \int \limits_{V^\alpha_0} (\mbN^T \mbu)_0 d\mbV_{\bxi} &= \sum \limits_{\beta = 0,1} (-1)^\beta
\left ( Q_{0,2}^{\alpha,\beta} - Q_{0,1}^{\alpha,\beta} \right ) +
\widehat{U}^\alpha \label{velocity_summary} \\
\int \limits_{V^\alpha_1} (\mbN^T \mbu)_1 d\mbV_{\bxi} &= \sum \limits_{\beta = 0,1} (-1)^\beta
\left ( Q_{0,1}^{\alpha,\beta} - Q_{1,2}^{\alpha,\beta} \right ), \label{radial_velocity}\\
\int \limits_{V^\alpha_2} (\mbN^T \mbu)_2 d\mbV_{\bxi} &= \sum \limits_{\beta = 0,1} (-1)^\beta
\left ( Q_{1,2}^{\alpha,\beta} - Q_{0,2}^{\alpha,\beta} \right ), \\
\int \limits_{V^\alpha_3} (\mbN^T \mbu)_3 d\mbV_{\bxi} &= 0,
\end{align}
\end{subequations}
where $Q_{0,1}$, $Q_{0,2}$, and $Q_{1,2}$ are defined by (\ref{Q01}),
(\ref{Q02}) and (\ref{Q12}), respectively, and
\begin{align}
\widehat{U}^{\alpha} & = \frac{2\pi}{m} \int_{\xi_1^0}^{\xi_1^1}
\int_{\xi_2^0}^{\xi_2^1} \mbB \cdot \left ( Z \mathbf{E} - \frac{\mu}{2}\nabla B \right) \left (\frac{\partial R}{\partial \xi_1} \frac{\partial Z}{\partial
  \xi_2} - \frac{\partial R}{\partial \xi_2} \frac{\partial
  Z}{\partial \xi_1} \right ) R d\xi_1 d\xi_2. \label{Udef}
\end{align}
We note that no metric factors appear in $Q_{0,1}$, $Q_{0,2}$, and
$Q_{1,2}$, which are computed exactly from the evaluation of magnetic field data
at configuration space cell vertices, except for
the cell face integrals of the transverse $\phi$ derivatives divided by $B$.  These
latter integrals can be computed using the fourth-order product
formula (\ref{productaverage}), in which the resulting $\phi$ integrals can be evaluated
exactly in terms of nodal $\phi$ values.  This fact plays a
critical role in avoiding a severe instability that can arise in
the high-order, finite-volume modeling of drift waves \cite{DoDoHiLeGh17}.

By tracing the individual terms of the gyrokinetic velocity
(\ref{gkvelocity}) through the
above development, the physical meaning of the terms in (\ref{Q01}),
(\ref{Q02}), (\ref{Q12}) and (\ref{Udef}) can be identified.  The
$\mathbf{B} \cdot \mathbf{E}$ and $\mathbf{B} \cdot \nabla B$
acceleration terms in (\ref{Udef}) are clear.  The first and second terms in
(\ref{Q01}) and (\ref{Q02}) are the $\mathbf{E} \times \mathbf{B}$ and
$\nabla B$ drifts, respectively.  The second term in
(\ref{Q12}) is the curvature drift.  The first term in
(\ref{Q12}) is the parallel streaming contribution, whose magnitude, as mentioned in
Section \ref{sec:gksystem}, dominates those of the drift terms by a few orders of
magnitude for typical tokamak parameters.  In addition to the fact
that the parallel streaming, $\nabla B$ drift and curvature drift terms
are all computed exactly, an important consequence
of the velocity discretization described here is that, on radial cell
faces $V^\alpha_1$ contained in a flux surface, parallel streaming
makes zero (to machine precision) contribution to the normal velocity
component (\ref{radial_velocity}), due to the subtraction of the
uniform $\Psi$ values effected by the $\beta$ sum.  This fact ensures that
parallel velocity discretization error cannot dominate the
drift terms, thereby masking the contribution of the latter via what
is sometimes referred to as ``numerical pollution'' \cite{UmanskyDayRognlien2005}.

\subsection{Multiblock extension \label{sec:mappedmultiblock}}

\begin{figure}
\centering
\includegraphics*[height=2.2in]{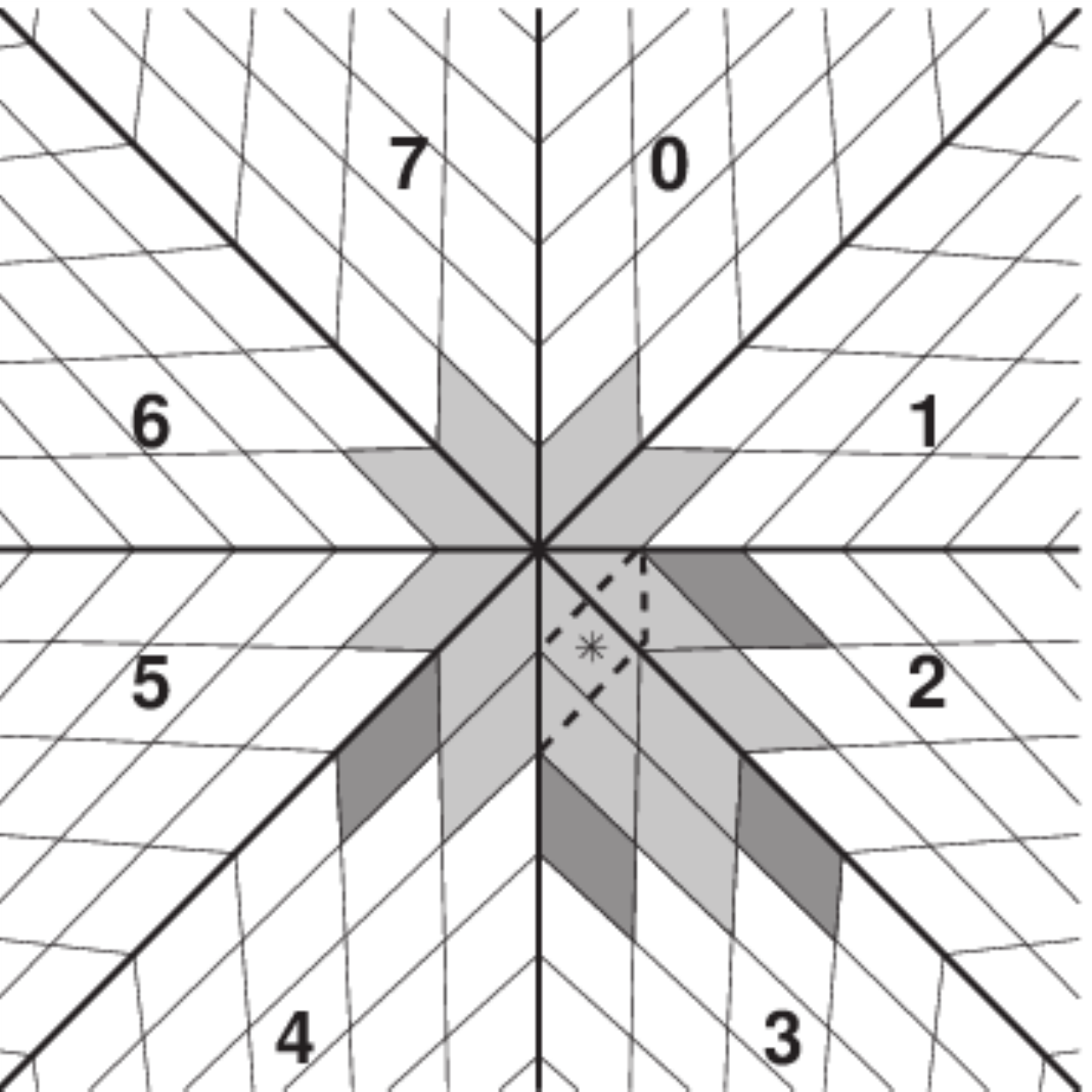}
\caption{Multiblock interpolation near the X point. The
    asterisk * indicates the center of an extrablock ghost cell of
    block number 4, which is filled by interpolation from the data in
    the shaded cells of neighboring blocks.  The shading distinguishes
a technical detail that is fully described in \cite{McCoDoHi2013}.} 
\label{fig:multiblockInterpolation}
\end{figure}

The single block discretization just described can be extended to the
multiblock edge plasma geometry (Figure \ref{fig:singlenull}) using
the approach detailed in \cite{McCoDoHi2013}.  In order to apply at
block interfaces the same fourth-order reconstruction
(\ref{fface}) used in the block interiors, a halo of ``extrablock''
ghost cells is filled by interpolation from surrounding cell-averaged
data.  Assuming that the smooth mappings on each block possess smooth
extensions beyond their respective boundaries, these extrablock ghost
cells are generated by simply applying the mapping to an extended
computational grid.  Figure \ref{fig:multiblockInterpolation} shows an
example of filling an extrablock ghost cell near the X point.  The
only aspect the calculation of fluxes at interblock boundaries that is
special is a post-processing step performed to restore strict
conservation.  Since there is no guarantee that the fluxes calculated
on each side of a multiblock interface using the above procedure will
agree to better than fourth order, the two fluxes are averaged to
define a consistent value for use on both adjacent blocks.

The key element is therefore the interpolation of valid neighbor block
data to extrablock ghost cells, which we briefly summarize here.
Full details are contained in \cite{McCoDoHi2013}.  We use a
least-squares approach that allows us to obtain high-order accuracy
independent of the degree of smoothness of the grid. We compute a
polynomial interpolant in the neighborhood of a ghost cell of the form
\begin{gather}
  \varphi(x) \approx \sum \limits_{p_d \geq 0 ; p_1 + \dots +
    p_D \leq P-1} a_{p} {x}^{p}
  \hbox{ , } p = (p_1 , \dots , p_D) \hbox{ , }
  {x}^{p} = x_1^{p_1} \dots x_D^{p_D}.
\end{gather}
We assume that we know the conserved quantities in a collection of
control volumes $\boldsymbol{v} \in \mathcal{V}$. In that case, we impose
the conditions 
\begin{gather}
  \int \limits_{v} \varphi(\xi) d\xi =  \sum \limits_{p_d \geq 0 ; p_1 + \dots +
    p_D \leq P-1} a_{p} \int \limits_{v} {x(\xi)}^{p} d \xi \hbox{ , }
  \boldsymbol \in \mathcal{V}. \label{leastsquares}
\end{gather} 
The integrals on the left-hand side can be computed to any order from
the known integrals of the conserved quantities $J \phi $, and the
integrals of ${x}^{p}$ can be computed
directly from the grid mapping.  Thus, this constitutes a system of
linear equations for the interpolation coefficients
$a_{p}$. Generally, we choose the number of equations to
be greater than the number of unknowns in such a way that the
resulting overdetermined system has maximal rank, so that it can be
solved using least squares. In the case where we are computing an
interpolant onto a finer grid from a coarser one in a locally-refined
grid calculation, we impose the conservation condition as a linear
constraint.

%% file: mapping.tex
\section{Mapping the edge geometry \label{sec:mapping}}

The use of mapped coordinates to accommodate strong anisotropy along
magnetic field lines requires the construction of block mappings from computational to physical
coordinates in which one of the computational coordinates
parameterizes flux surfaces.  In this section, we describe an
approach based on the assumed availability of 
the magnetic flux $\Psi = \Psi(R,Z)$ in cylindrical coordinates, from which the (assumed
axisymmetric) magnetic field $\mbB = (B_R, B_\Phi, B_Z)$ is
obtained as
\begin{equation}
B_R(R,Z) = - \frac{1}{R}\frac{\partial \Psi(R,Z)}{\partial Z},
~~ B_\Phi(R,Z) = \frac{(RB)_{\rm tor}}{R}, ~~B_Z(R,Z) =
\frac{1}{R}\frac{\partial \Psi(R,Z)}{\partial R}, \label{field_def}
\end{equation}
where $(RB)_{\rm tor}$ is a constant.  This provides a smooth
representation of the magnetic field for use in the construction of
mappings as well as the evaluation of the gyrokinetic velocity
(\ref{gkvelocity}).  The magnetic
flux may be obtained, for example, as the result of a separate equilibrium calculation
or a spectral interpolation of experimental flux measurements.

\begin{figure}
\centering
\begin{picture}(600,220)
\put(70,-20){\includegraphics[height=3.5in]{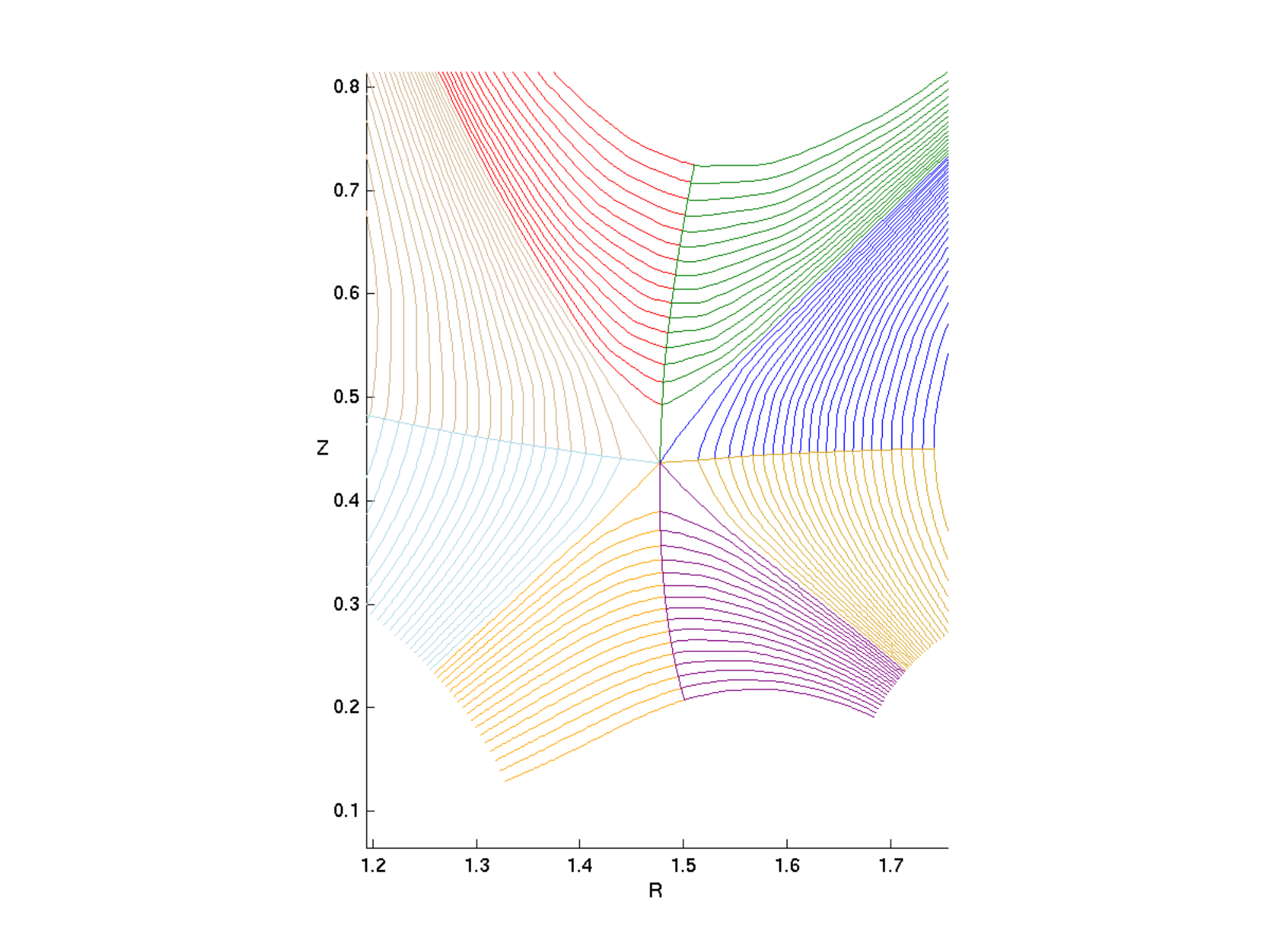}}
\end{picture}
\caption{Flux surfaces in the vicinity of the X point become
  increasingly kinked, precluding the introduction of a smooth
  flux-aligned coordinate system in this area.\label{fig:xpt_vicinity}} 
\end{figure}

Although field lines define smooth curves in three-dimensional space,
for axisymmetric edge geometries the use of fully flux-aligned
coordinate mappings is problematic near the X point.  As shown in
Figure \ref{fig:xpt_vicinity}, flux surfaces become increasingly
``kinked'' approaching the X point, resulting in diverging metric
factors for flux-aligned coordinate mappings.  As noted at the end of
Section \ref{sec:divfree}, no metric factors appear in the normal
velocity integrals (\ref{velocity_summary}), so the singularity of
flux-aligned metrics does not affect their calculation.  On the other
hand, reconstruction of the distribution function on block interfaces
using the interpolation described in Section
\ref{sec:mappedmultiblock} requires not only smooth mappings up to
block boundaries, but also some distance beyond to accommodate the
ghost cell halo.  We therefore relax the assumption of flux-aligned
mappings near the X point.  Although the cost of such a modification
is likely the need for increased resolution, this is mitigated by the
fact that, since the poloidal component of the magnetic field is small
near the X point, anisotropy is less of a concern there and thus the
need for strict flux alignment is reduced.

The poloidal grid lines of a dealigned grid can be constructed as the level
surfaces of a modified flux function $\Psi$ obtained by
the blending of the original magnetic flux function $\Psi_0$, given in
the vicinity of the X point by
\begin{equation}
\Psi_0 - \Psi_X \approx \bar{R}^2 - \bar{Z}^2,   \label{Xhyperbola}
\end{equation}
and the block-aligned linear function
\begin{equation}
\Psi_{\rm lin} = D \left ( |\bar{R}| - |\bar{Z}| \right ), \label{psilindef}
\end{equation}
such that outside a transition radial distance $D$, the blended flux
becomes the original flux, whereas inside the flux is given by the
block-aligned flux.  Here, $\Psi_X$ is the value of $\Psi_0$ at the X
point $(R_X,Z_X)$ and $(\bar{R},\bar{Z})$ are the linear coordinate
transforms of $(R,Z)$. We define the blended flux function by
\begin{align}
\Psi = \Psi_X + \tanh \left ( \frac{r}{D} \right ) \left ( \Psi_0
- \Psi_X \right ) + \alpha \left ( 1 - \tanh \left ( \frac{r}{D}
\right ) \right ) \Psi_{\rm lin}, \label{hybridflux}
\end{align}
where $r = \sqrt{\bar{R}^2 +
\bar{Z}^2}$, and $\alpha$ is an optimization parameter of the order unity.

The linear coordinate transforms $(\bar{R},\bar{Z})$ are related to
the $(R,Z)$ coordinate system by
\begin{subequations}
\begin{align}
\bar{R} &= a_1 (R - R_X) + b_1 (Z - Z_X), \\
\bar{Z} &= a_2 (R - R_X) + b_2 (Z - Z_X),
\end{align}
\end{subequations}
where the linear coefficients can be found by solving a system of
nonlinear algebraic equations.  Introducing
the original flux function expansion near the X point as 
\begin{align}
\Psi_0 - \Psi_X = a(R - R_X)^2 + b(R - R_X)(Z - Z_X) + c(Z - Z_X)^2 \label{quadform}
\end{align}
and making use of (\ref{Xhyperbola}), we obtain
\begin{subequations}
\begin{align}
a_1^2 - a_2^2 &= a, \label {matching}\\
\nonumber b_1^2 - b_2^2 &= c, \\
\nonumber 2 a_1 b_1 - 2 a_2 b_2 &= b, \\
\nonumber a_1 a_2 + b_1 b_2 &= 0,
\end{align}
\end{subequations}
where the last equation in (\ref{matching}) is the orthogonality
condition of the $(\bar{R},\bar{Z})$ coordinate system.  Because the
quadratic form (\ref{quadform}) for a magnetic flux near the X point
is hyperbolic, $b^2 - 4ac > 0$, which implies that $b^2 + 2c (-a \sqrt{b^2
  + (a-c)^2} + c) > 0$ if $b \ne 0$ (see Appendix \ref{sec:inequality}).
Letting $\delta = \sqrt{b^2 + (a-c)^2}$, the solution
of the system (\ref{matching}) is then
\begin{subequations}
\label{mapcoefs}
\begin{align}
a_1 &= - \frac{
\left ( a + \delta - c \right ) \sqrt{ b^2 +2c \left ( -a + \delta + c \right ) }
}{
2b \delta^{1/2}
}, \\
a_2  &= - \frac{
\sqrt{2a^2 + b^2 -2a \left ( \delta + c \right )}
}{
2 \delta^{1/2}
}, \\
b_1 &= - \frac{
\sqrt{b^2 +2c \left ( -a + \delta + c \right )}
}{
2 \delta^{1/2} 
}, \\
b_2 &= - \frac{
\left ( a + \delta - c \right ) \sqrt{ 2a^2 +b^2 - 2a \left (
  \delta + c \right ) }
}{
2b \delta^{1/2}.
}
\end{align}
\end{subequations}
Solutions for the special case of $b = 0$ can be obtained by applying
L'Hospital's rule to the above expressions.

Given the hybrid flux (\ref{hybridflux}) just constructed, the desired
poloidal grid lines correspond to level surfaces of $\Psi$ spaced
equidistantly in $\Psi$ and are obtained by ray tracing.  The radial grid lines are obtained as follows:
\begin{enumerate}
\item Construct an arc length mapping
  $(R(\ell_\theta),Z(\ell_\theta))$ along the (modified) separatrix;
\item Position grid nodes on the separatrix such that they are
  equidistant in $\ell_\theta$; then
\item From those points, ray trace radial grid lines as follows.  In the
  MCORE and MCSOL blocks, trace the radial grid lines normal to
  $\Psi$, yielding a locally orthogonal grid in those blocks.
  In the remaining blocks, trace the radial
  grid lines parallel in directions given by a smooth blending of the
  normal-to-$\Psi$ and block boundary directions.
The purpose of this blending is
to ensure that ({\em i\/}) the grid is locally orthogonal in a
neighborhood of such radial block boundaries and ({\em ii\/}), in a
neighborhood of the radial block boundaries that do contain the X
point, the radial grid lines are nearly parallel to those boundaries.
The purpose of goal ({\em i\/}) is to enable the construction of a
global MMB grid in which ghost cells overlap the valid cells of
neighboring blocks at the poloidal interface between MCORE and LCORE
or RCORE, as well as between MCSOL and LCSOL or RCSOL, so that ghost
cell data can be obtained by direct copying rather than
interpolation.  Goal ({\em ii\/}) avoids the generation of small cells
that can result in an unnecessary stability limitation for explicit
time integration, as well as the possibility of large derivatives in
the metric factors computed from the resulting mapping.

\end{enumerate}

\begin{figure}
\centering
\begin{picture}(600,220)
\put(-80,-140){\includegraphics[width=5.5in]{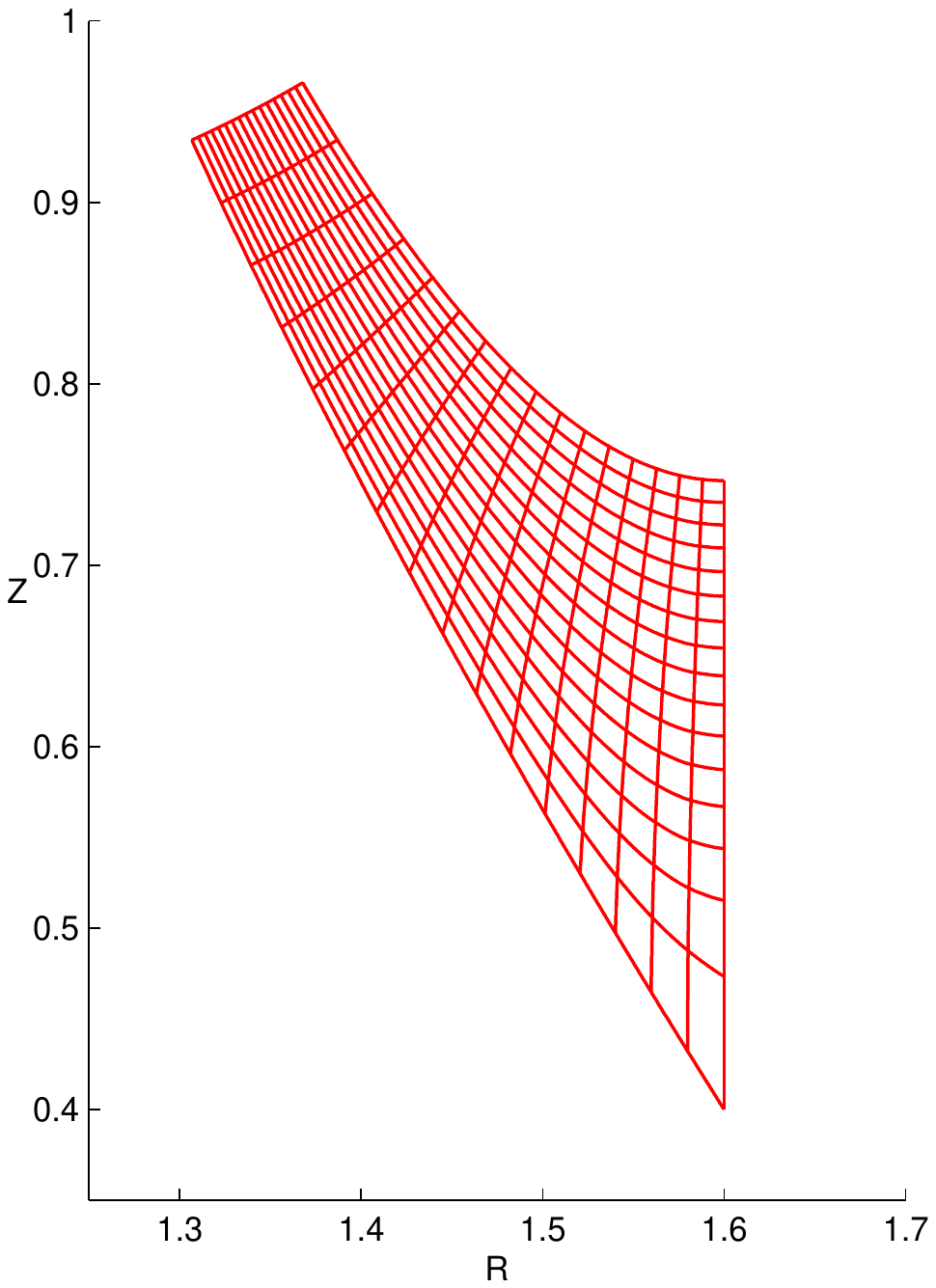}}
\put(120,-140){\includegraphics[width=5.5in]{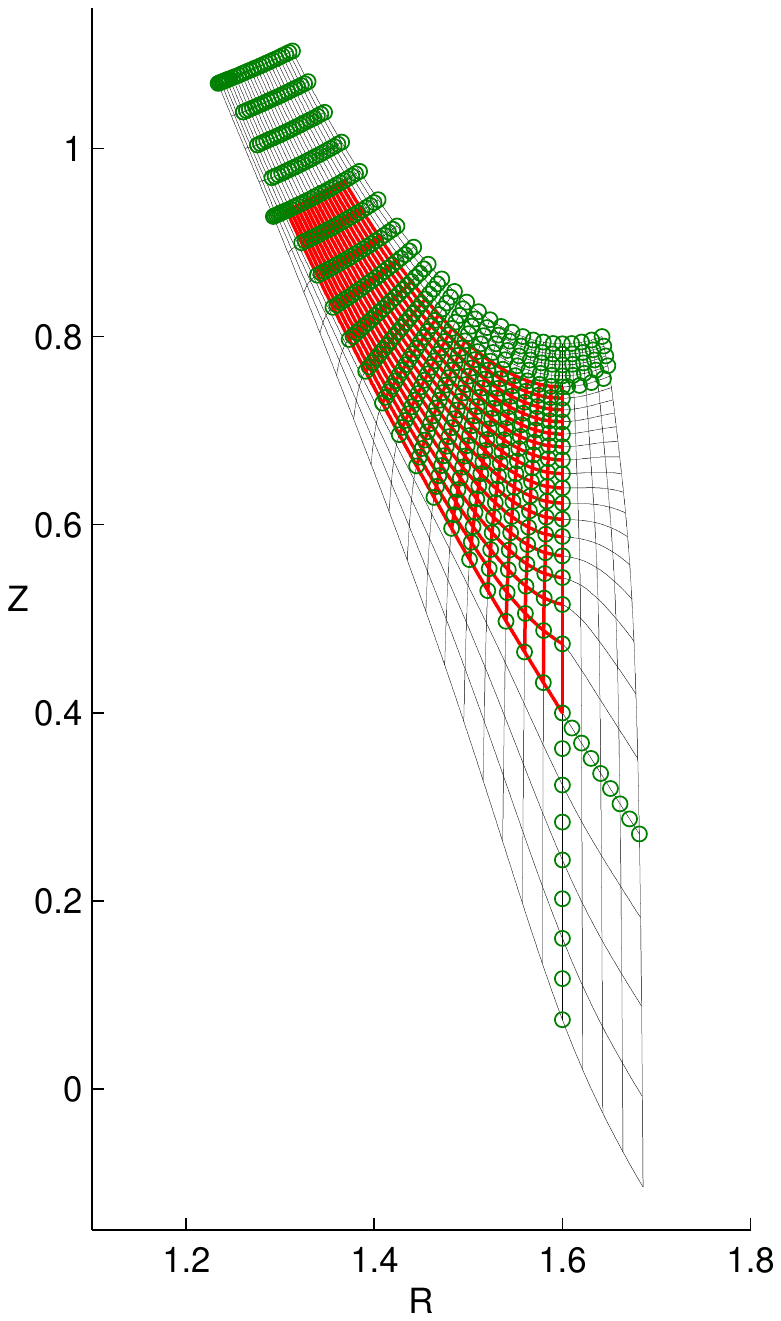}}
\put(112,22){X point $\rightarrow$}
\end{picture}
\caption{Construction of the LCORE mapping grid.  Left: A
  grid is generated that is flux-dealigned near the X point,
  boundary-aligned near the right-hand boundary and locally-orthogonal away from the X point.  Right: Ghost cells are added by
  evaluating RBF interpolants of the ($R,Z$) coordinates of the points indicated by the circles,
  comprised of the union of the valid block grid (red) points, physical boundary points and
  points along the extended block boundaries. \label{fig:dealign}}
\end{figure}

\noindent Figure \ref{fig:dealign} (left) shows the result (the red grid) of
applying the above procedure in the LCORE block of the magnetic
equilibrium used in the numerical tests performed in Section
\ref{sec:verification}.  The black grid is the original flux-aligned grid
for which poloidal grid lines are obtained as level surfaces of the
magnetic flux function.

To add ghost cells, the block boundaries of the dealigned block grid are first extended in each
direction the desired number $M$ of ghost cells.  In the LCORE example
shown on the right-hand side of Figure \ref{fig:dealign}, $M = 4$ points
(indicated by circles) have been added along each boundary extension.  These
points (32 in the Figure \ref{fig:dealign} example) are combined with the
block grid points and a layer of $M$ points along the two block
boundaries not containing the X point (which are obtainable from
the magnetic flux in the same manner as the aligned interior points) to obtain a set of points $\{
(R(\xi_i,\eta_i),Z(\xi_i,\eta_i)): 1 \le i \le N\}$ representing $N$
evaluations of a mapping $(\xi,\eta) \rightarrow
(R(\xi,\eta),Z(\xi,\eta))$ from logical to physical coordinates.
We obtain the desired mapping as the radial basis function (RBF) interpolants 
\begin{subequations}
\label{rbf_interp}
\begin{align}
R (\xi,\eta) &=\ \sum_{i=1}^N \alpha_{R,i} [ (\xi - 
\xi_{i})^2 + (\eta - \eta_{i})^2  ]^{3/2},  \\
Z (\xi,\eta) &=\ \sum_{i=1}^N \alpha_{Z,i} [ (\xi - 
\xi_{i})^2 + (\eta - \eta_{i})^2  ]^{3/2},
\end{align}
\end{subequations}
by solving linear systems for the coefficients $(\alpha_{R,i},\alpha_{Z,i})$ such that
(\ref{rbf_interp}) interpolates the points
$(R(\xi_i,\eta_i),Z(\xi_i,\eta_i)), 1 \le i \le N$.  The remaining
ghost cell vertex coordinates are then obtained by evaluating
(\ref{rbf_interp}) at the corresponding logical indices, yielding the
full ghost cell region, as indicated by the black grid in the right-hand
side of Figure \ref{fig:dealign}.  

Since the MCORE and MCSOL blocks do not contain the X point,
flux-aligned locally orthogonal grids in these blocks can be generated
by simple ray tracing.  Because the grid dealignment in the
blocks that do contain the X point is limited to a region that is
bounded away from the MCORE and MCSOL block boundaries due to the
the ray tracing of radial grid lines (see Section \ref{sec:mapping}), ghost cells at all
intrablock boundaries involving MCORE and MCSOL can be obtained from
valid cells in adjacent blocks.  For ghost cells constructed in this manner, the
filling of the ghost cell halo described in Subsection
\ref{sec:mappedmultiblock} can be performed by direct copying of valid
data rather than interpolation.

\begin{figure}
\centering
\setlength{\unitlength}{0.01in}
\begin{picture}(600,370)
\put(-220,-200){\includegraphics[width=6in]{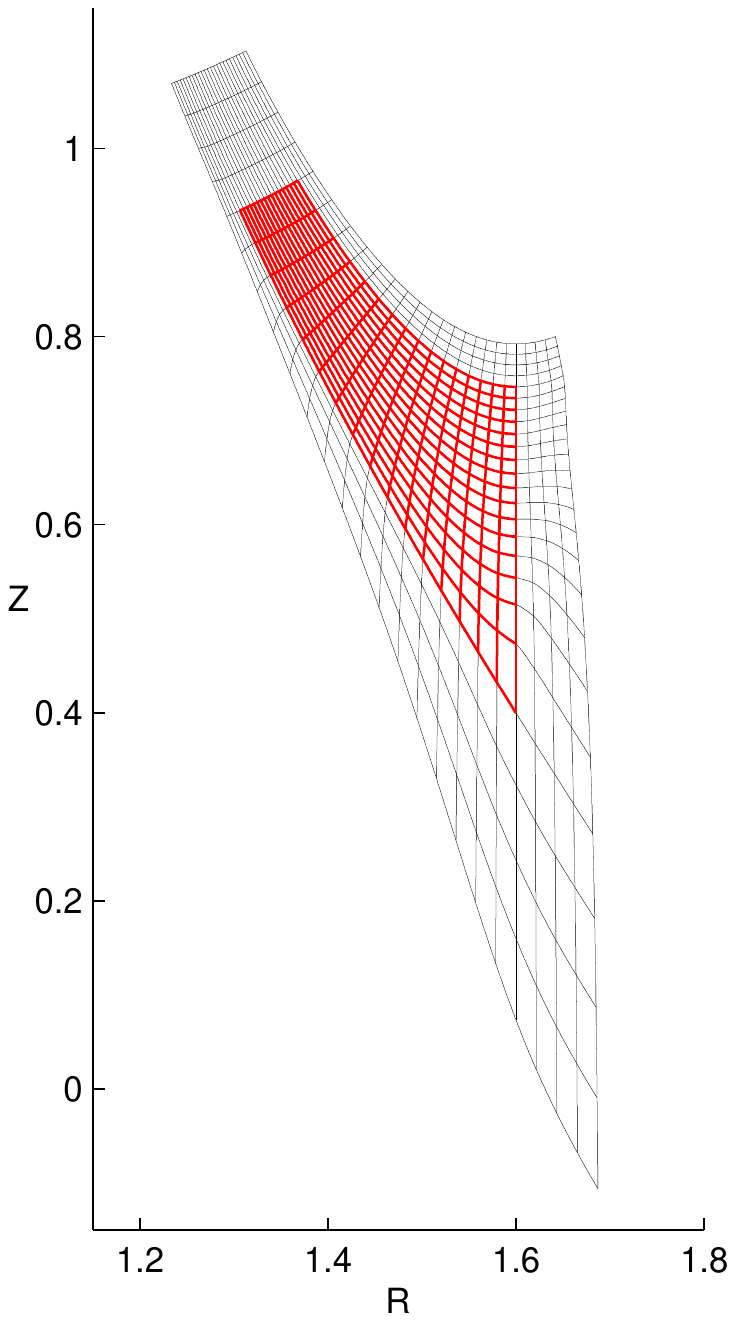}}
\put(100,-200){\includegraphics[width=6in]{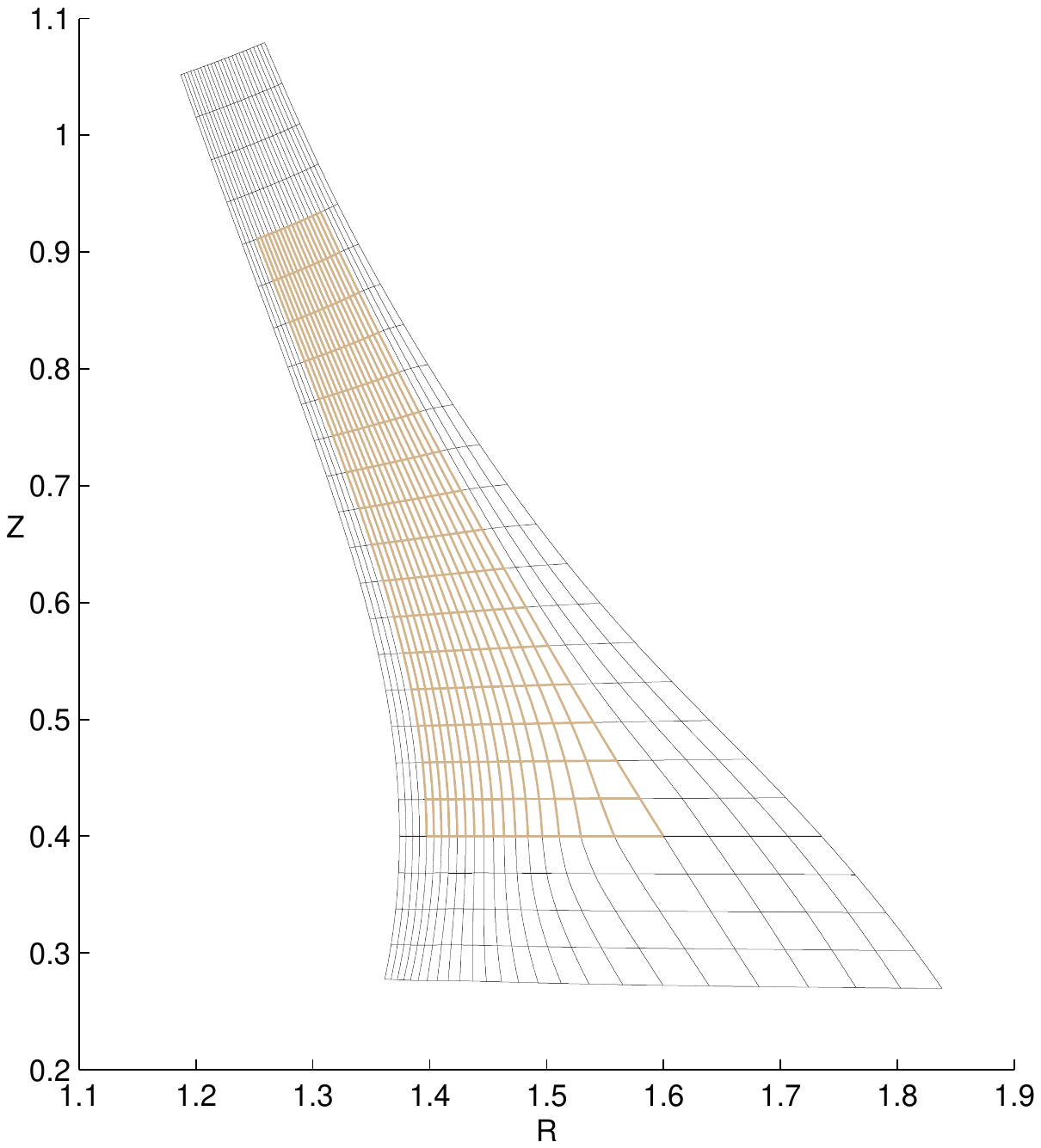}}
\end{picture}
\caption{LCORE (left) and LCSOL (right) mapping grids.\label{lcore_lcsol_block}} 
\end{figure}

\begin{figure}
\centering
\setlength{\unitlength}{0.01in}
\begin{picture}(600,370)
\put(-160,-200){\includegraphics[width=6in]{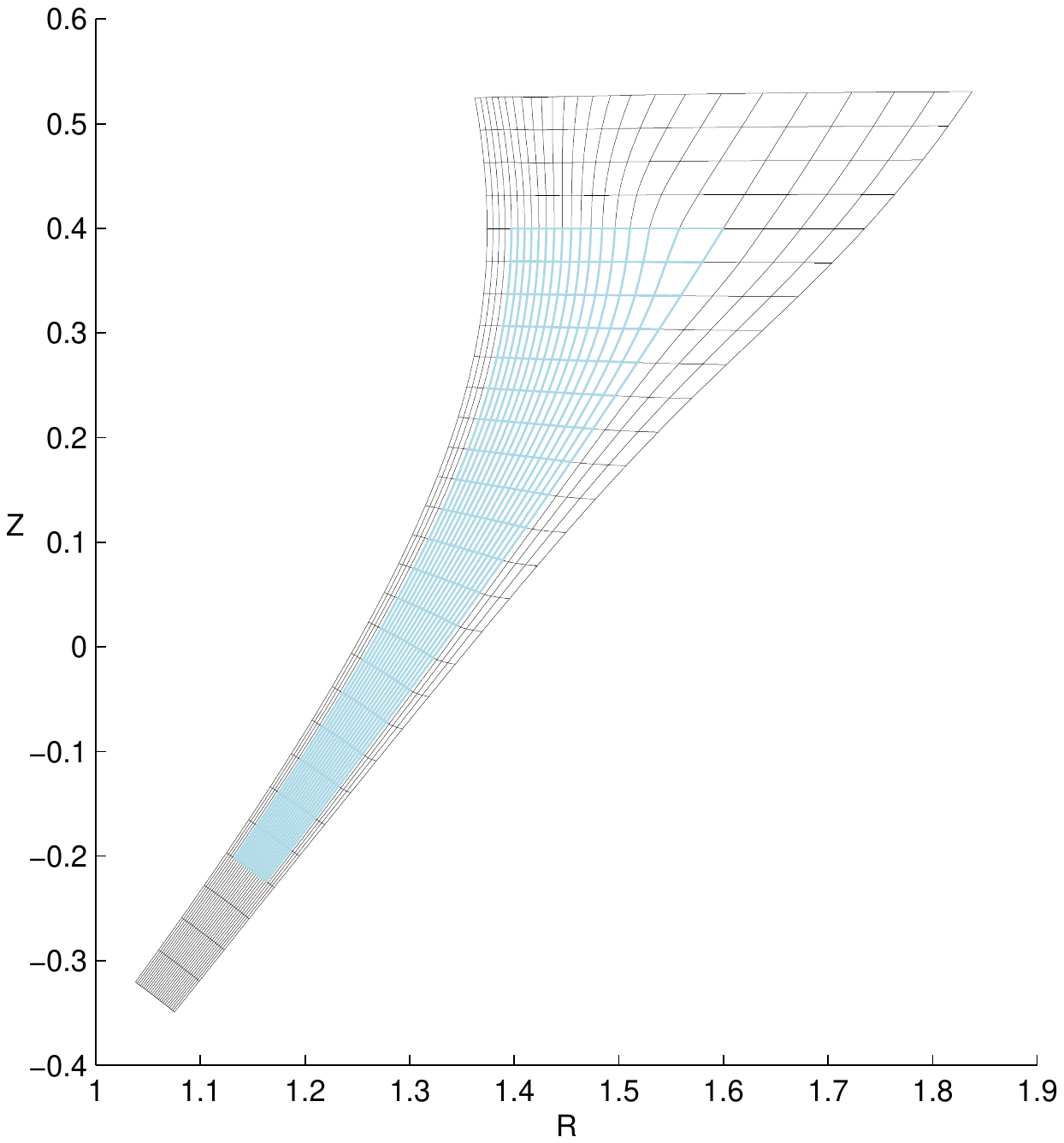}}
\put(150,-200){\includegraphics[width=6in]{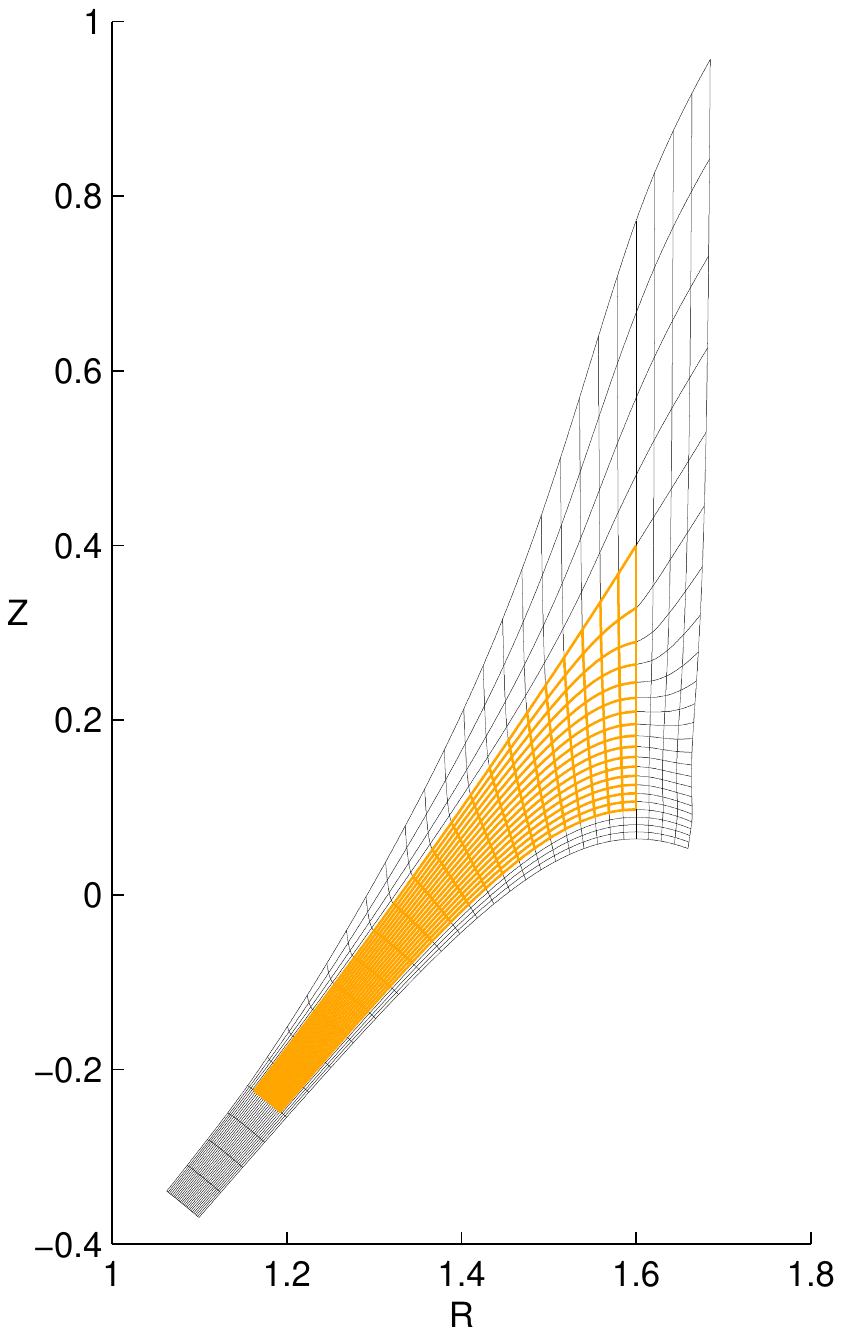}}
\end{picture}
\caption{LSOL (left) and LPF (right) mapping grids.\label{lsol_lpf_block}} 
\end{figure}

\begin{figure}
\centering
\setlength{\unitlength}{0.01in}
\begin{picture}(600,370)
\put(-160,-200){\includegraphics[width=6in]{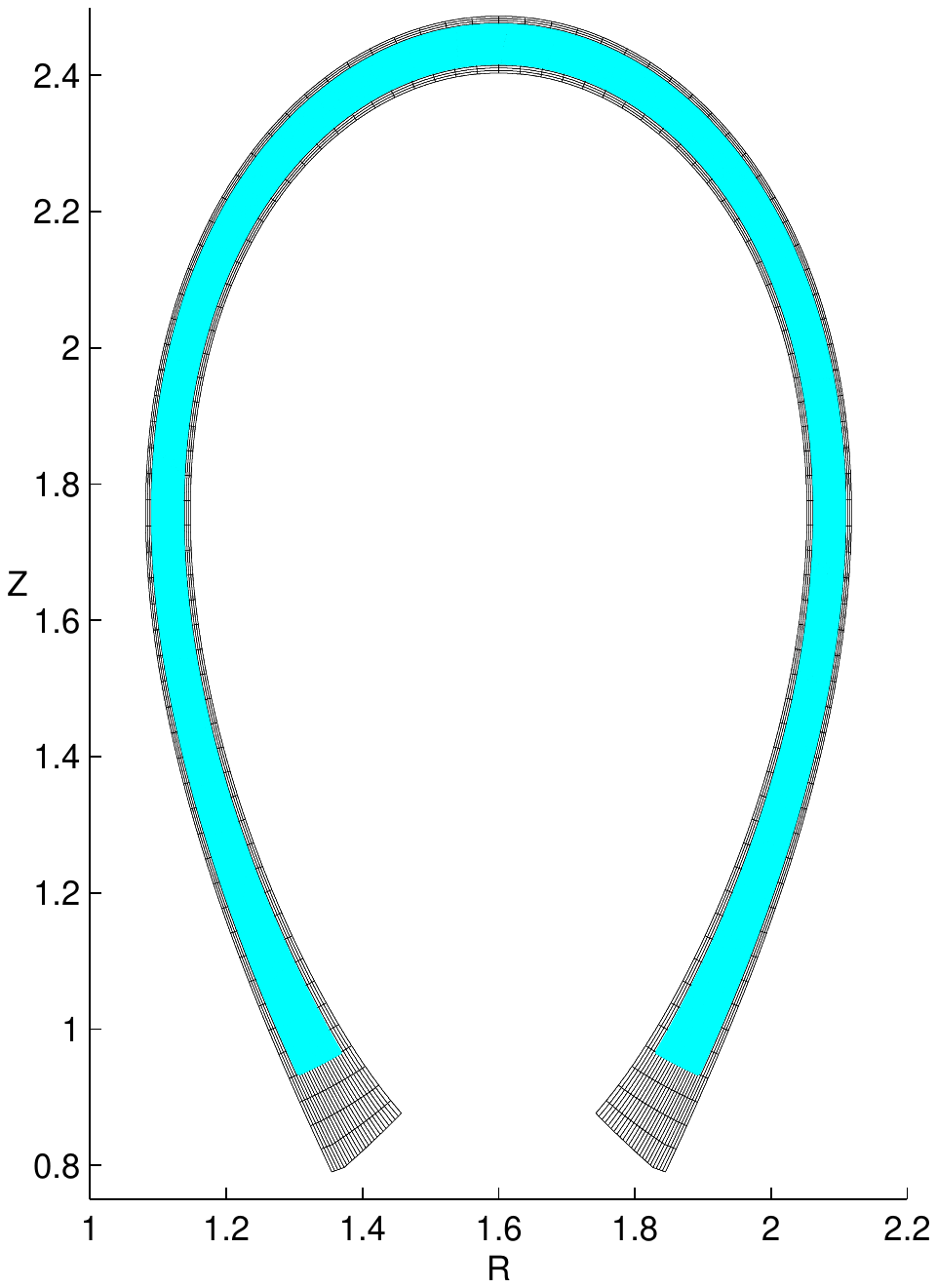}}
\put(150,-200){\includegraphics[width=6in]{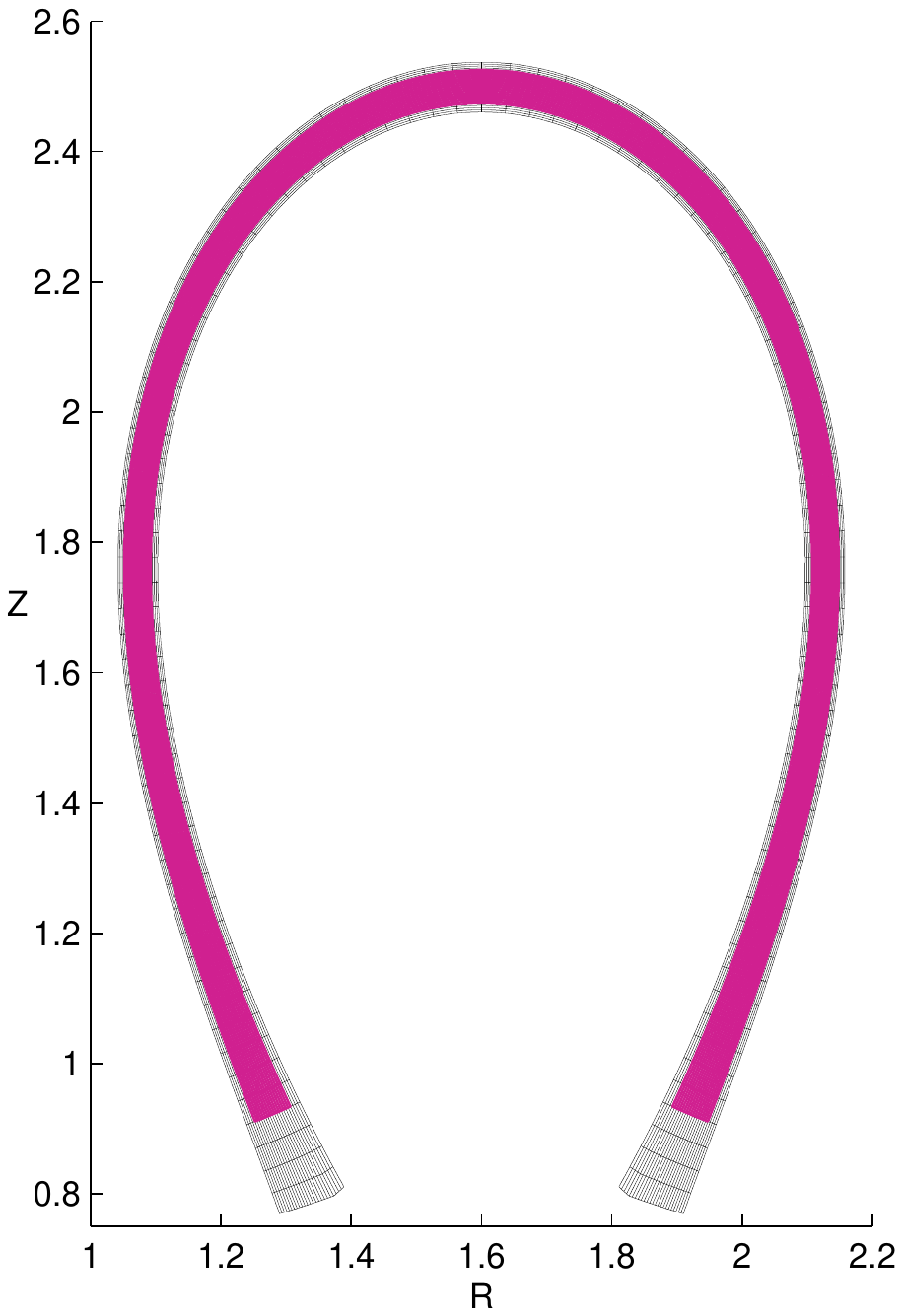}}
\end{picture}
\caption{MCORE (left) and MCSOL (right) mapping grids.\label{mcore_mcsol_block}} 
\end{figure}

Figures \ref{lcore_lcsol_block} through \ref{mcore_mcsol_block} show
the extended block grids generated by the above approach for the
LCORE, LCSOL, LSOL, LPF, MCORE and MCSOL blocks of the Section
\ref{sec:verification} example, coarsened by a factor of two for
better plotting visibility.  The RCORE, RCSOL, RSOL and RPF
blocks are symmetric with their left-hand side counterparts.  The
black grid lines demarcate the extended ghost cells.

The purpose of the grid generation just described is to enable the
discrete representation of the block mappings required by the mapped
multiblock discretization described in Section \ref{sec:discretization}.
Using the extended block grids ({\em e.g.},
Figures \ref{lcore_lcsol_block} through \ref{mcore_mcsol_block}), we
construct sixth-order B-spline interpolants of the $(R,Z)$ physical
coordinate components, which yields mappings with four continuous derivatives.
The spatial grid used to discretized the Vlasov system is then the
image under this mapping of a computational grid at the desired
resolution, which in general will not be the same as the block grids
used to define the mapping.  The grid generation performed to define the
mapping is only performed once for a given magnetic flux.  A few additional
points are in order concerning this approach:
\begin{enumerate}
\item The RBFs used in (\ref{rbf_interp}) are also known as polyharmonic
  smoothing splines \cite{PolyharmonicSplines}.  In addition to being smooth, such functions
  minimize a particular functional of higher derivatives, which makes
  them a robust choice for the high-order extrapolation being
  performed here.
\item The RBF interpolants (\ref{rbf_interp}) might also appear to provide
  a convenient mapping choice, having already 
  constructed them for the purpose of generating ghost cells.
  Unfortunately, since each such block interpolant is defined using
  data specific to its own block, there is no guarantee that the
  interpolants will be continuous along interblock boundaries ({\em
    i.e.}, {\em between} common grid points where they do in fact
  agree), which is required by our mapped discretization approach.  In contrast,
  due to their tensor product definition, the B-spline
  interpolants are uniquely defined by their values at the shared
  points along interblock boundaries and are therefore continuous
  across such boundaries.
\item Although the vertices of the block grids used to define the
  block mappings are flux-aligned by construction, except in the
  dealignment neighborhood of the X point, the vertices of grids
  obtained by evaluating the B-spline interpolants are not guaranteed to
  be similarly flux-aligned.  It is therefore important to employ an adequate number of
  poloidal points in the block mapping grids to
  reduce the error due to grid dealignment.  In studies we
  have performed to date, we find that using 128 or 256 poloidal cells
  in the core region of the mapping grid is
  likely sufficient for realistic edge geometries.  If dealignment
  error is an acute concern for some reason, one always has the option of
  employing a grid consistent with that used to construct the mapping
  so that no interpolation is performed. 
\end{enumerate}

%% file: cogent.tex
\section{COGENT \label{sec:cogent}}
The spatial discretization approach described above provides the foundation of the
COGENT (\underline{CO}ntinuum \underline{G}yrokinetic \underline{E}dge
\underline{N}ew \underline{T}echnology) code
\cite{DoEtAl2012,CoEtAl2013,DoCoCoDoHi10,DorfEtAl2013,DorfEtAl2016},
which we have been developing for the solution of gyrokinetic systems
in mapped multiblock geometries, including those describing the
tokamak edge.  Although this paper is addressing just the Vlasov operator
discretization, we include this short description of the more general
code environment in which it is being used.

To obtain self-consistent electric fields, COGENT includes an
electrostatic model.  The potential $\Phi$ depends upon the
charge density of the species distribution functions $f_\alpha$ being evolved
by (\ref{conservativenormalized})-(\ref{gkvelocity_vars}), which in turn depend
upon $\Phi$ in the velocity calculation.  Because the
$f_\alpha$ are computed in gyrocenter coordinates and the Poisson
equation is posed in the lab frame, the velocity integral yielding
the ion charge density must therefore be split into two
pieces.  In the long wavelength limit $k_\perp \rho \ll 1$, where $k_\perp$ is the
magnetic field perpendicular wave number and $\rho$ is the ion gyroradius, the
gyrokinetic Poisson equation is
\begin{equation}
 {\bnabla}_{\mbX} \cdot \left \{ \left [
  \lambda_D^2  \mbI + \rho_L^2
  \sum_{\alpha} \frac{Z_{\alpha}
  \bar{n}_{\alpha}}{m_{\alpha} \Omega_{\alpha}^2} \left ( \mbI -
  \mbb\mbb^T \right ) \right ]  {\bnabla}_{\mbX} \Phi \right \} =
  n_e - \sum_{\alpha} Z_{\alpha} \bar{n}_{\alpha} ,  \label{gkpoisson}
\end{equation}
where $ {\bnabla}_{\mbX}$ denotes the gradient with respect to
the normalized lab frame coordinate and $\lambda_D$ is the normalized
Debye length.  The quantity
\begin{equation}
\bar{n}_\alpha(x,t) =
\frac{1}{m_\alpha} \int f_\alpha(x, v_\|, \mu, t) 
  B_{\parallel}^*(x,\vpll) d\vpll d{\mu}  \label{nidef}
\end{equation}
is the ion gyrocenter density, which is the gyrophase-independent part
of the integration of the gyrocenter distribution function $f_\alpha$
over velocity.  The second term in the left-hand side of
(\ref{gkpoisson}) is the polarization density, which is the
gyrophase-dependent part of the velocity integration of $f_\alpha$.
Since this piece depends upon the potential, we must combine it with
the usual Laplacian (the first term in (\ref{gkpoisson})) in the
construction of the linear operator to be solved for $\Phi$.  Here,
$\mbb$ denotes the unit vector in the direction of the magnetic field,
$Z_\alpha$ is the charge state, $m_\alpha$ is the mass, and
$\Omega_\alpha$ is the gyrofrequency.  We note that for typical
tokamak parameters, $\lambda_D \ll \rho_L$, and hence the polarization
density term dominates.  Because the electron gyroradius is small, a
similar splitting of the electron density is omitted.

COGENT includes a range of collision operators, which appear as
source terms in the right-hand side of (\ref{conservativenormalized}).
These include a Krook model, a drag-diffusion operator in parallel
velocity space, Lorentz collisions, a linearized Fokker-Planck model
conserving momentum and energy \cite{DoEtAl2012}, and a fully
nonlinear Fokker-Planck model of Coulomb collisions
\cite{DorfEtAl2014}.

Time integration in COGENT is performed using either an explicit
fourth-order Runge Kutta (RK4) method or a semi-implicit additive Runge-Kutta (ARK) time
integration scheme, via second-order (three-stage)
\cite{giraldokellyconsta2013}, third-order (four-stage)
\cite{kennedycarpenter}, or fourth-order (six-stage)
\cite{kennedycarpenter} options.
The ARK approach is  used to treat fast
collisional time scales and/or electron dynamics.  In both methods,
evaluation of the discrete Vlasov operator as described herein is
performed as a function evaluation, where the gyrokinetic Poisson
equation (\ref{gkpoisson}) is solved at each stage using the predicted distribution
functions, yielding the self-consistent electric field needed to compute the phase
space velocity (\ref{gkvelocity}).  In the
ARK approach, a Jacobian-free Newton-Krylov method is used to solve the
nonlinear equations that arise in the implicit updates \cite{Ghosh2017}.

COGENT is built upon the Chombo structured adaptive mesh refinement
(AMR) framework \cite{Chombo}.  Although COGENT does not currently
utilize Chombo's AMR capabilities, Chombo provides many data structures
and parallel operations that are well-suited for mapped multiblock
algorithms, including data containers for mesh-dependent quantities
distributed over processors and support for the mapped grid formalism
described in Section \ref{sec:discretization}.  Multidimensional
data types and operations, such as the computation of two-dimensional
densities from four-dimensional distribution functions and the
ability to independently decompose configuration and phase space, are
also provided.

%% file: verification.tex
\section{Truncation error verification \label{sec:verification}}

In this section, we demonstrate the high-order convergence of the
strategy described in the preceding sections when applied to the spatial
discretization of (\ref{conservativenormalized})-(\ref{gkvelocity_vars}) in an edge geometry.
Let
\begin{equation}
f(\psi,\theta,\vpll,\mu) = \frac{n(\psi)}{\pi^{1/2} (2T/m)^{3/2}} \exp \left (
- (m \vpll^2 + \mu B(\psi,\theta)) / 2T \right )  \label{maxwellian_ic}
\end{equation}
be a Maxwellian distribution function with density
\begin{equation}
n(\psi) = \tanh\left ( 25(0.9 - \psi) \right ) + 1.1
\end{equation}
and temperature $T = 1$.  Also let
\begin{align}
\phi = - \frac{T}{Z} \ln(n).   \label{potential_ic}
\end{align}
The above expressions incorporate the normalizations described
in Appendix \ref{normalizations}.  As shown in Appendix \ref{boltzmann_eq}, the Vlasov operator in
(\ref{conservativenormalized})-(\ref{gkvelocity_vars}) with the electric field computed from
(\ref{potential_ic}) vanishes when applied to the Boltzmann
equilibrium distribution function (\ref{maxwellian_ic}).  The application of the discretized operator therefore
represents the truncation error of the mapped multiblock spatial
discretization approach, which we can measure directly.  In obtaining the
results reported below, we employ the fourth-order,
centered-difference formula (\ref{fface}) with WENO-like limiter
modifications described in \cite{HittingerBanks2013}.

We analytically prescribe a magnetic flux defined in
\cite{DorfEtAl2016}, which results in a magnetic geometry
that is roughly characteristic of the DIII-D
tokamak.  Although our approach only requires a smooth flux
representation, including one obtained by interpolation from
experimental data, we consider an analytic model here to avoid any
additional complications from noise or other observational errors.
Specifically, we consider a normalized poloidal flux function
\begin{align}
\Psi_N(R,Z) = \cos \left [ c_1(R-R_0)/L_N \right ] + c_2\sin
\left [ (Z-Z_0)/L_N \right ] - c_3(Z-Z_0)/L_N,
\end{align}
where $R$ and $Z$ are the radial (distance from the magnetic axis) and
vertical (parallel to the magnetic axis) coordinates, respectively;
$Z_X = - L_N \arccos(c_3/c_2)$ corresponds to the vertical
position of the X point; $c_1 = 1.2$, $c_2 = 0.9$, and $c_3 = 0.7$ are
the constant shape factors; $L_N = 1{\rm (m)}$ is a normalizing spatial scale;
$R_0 = 1.6{\rm (m)}$ is the major radius coordinate corresponding to the
location of the magnetic axis; and $Z_0 = 0.4{\rm (m)} + L_N
\arccos(c_3/c_2)$ is a constant vertical shift adopted for
visualization purposes.  For the simulations reported, the radial
width of the open and closed field line regions is taken to be
$\Delta_R = 6.7{\rm (cm)}$ as measured at the top of the tokamak.  The
magnetic field is computed from (\ref{field_def}), where $(RB)_{\rm tor} =
3.5$(T-m) and where the poloidal flux is computed from
\begin{align}
\Psi = \Psi_N \bar{B}_\theta R_{\rm mp} \left [
   \left ( \frac{\partial \Psi_N}{\partial R} \right )^2 + \left ( \frac{\partial
        \Psi_N}{\partial Z} \right )^2 \right ]_{(R_{\rm mp},Z_{\rm mp})}^{-1/2}.
\end{align}
Here, $\bar{B}_\theta = 0.16{\rm (T)}$ is the magnitude of the poloidal
magnetic field at the intersection of the separatrix and the outboard
midplane corresponding to $R_{\rm mp} = 2.11{\rm (m)}$ and $Z_{\rm mp} = Z_0 + L_N
\arccos(c_3/c_2) = 1.76{\rm (m)}$, and the directions of the coordinate system
unit vectors are such that $\mathbf{e}_R \times \mathbf{e}_\phi = \mathbf{e}_Z$.
Using this analytically defined flux, the grid generation procedure
described in Section \ref{sec:mapping} yields the mapping blocks
displayed in Figures \ref{lcore_lcsol_block} through
\ref{mcore_mcsol_block}.

We consider the four-grid sequence specified in Table \ref{tab:grids},
obtained by refining each phase space dimension by a factor of two
relative to an initial grid.  We are interested in the behavior of the
truncation error at locations were the error is likely to be greatest,
rather than some globally integrated measure.  We therefore inspect
the error at block corners and boundaries, both physical and
intrablock, as well as a few points in the interior of blocks to
quantify the performance of the interior scheme there as well.  Since
the phase space geometry is multiblock only in the configuration space
coordinates, we also want to ensure that the error is not dominated by
the velocity space discretization, and thus possibly masking error due
to the multiblock treatment of configuration space.  We therefore
truncate the velocity space domain to $-1 \le \vpll \le 1$ and $0 \le
\mu \le 2$, relative to the thermal velocity scaling described in
Appendix \ref{normalizations}, which avoids the need for even finer
velocity space grids than those shown in Table \ref{tab:grids}, which
we have verified are sufficient to achieve adequate resolution.

\begin{table}
\center
\begin{tabular}{|c|c|c|c|c|r|} \hline
Grid & MCORE & Blocks containing & $\vpll$ & $\mu$ & Number of cells
\\ 
$m$  & MCSOL & the X point &         &       &\\
\hline \hline
$1$ & $8 \times 48$   & $8 \times 8$   & $24$  & $24$ & 737,280 \\ \hline
$2$ & $16 \times 96$  & $16 \times 16$ & $48$  & $48$ & 11,796,480 \\ \hline
$3$ & $32 \times 192$ & $32 \times 32$ & $96$  & $96$ & 188,743,680\\ \hline
$4$ & $64 \times 384$ & $64 \times 64$ & $192$ & $192$ & 3,019,898,880 \\ \hline
\end{tabular}
\caption{Grid sizes for the refinement study. \label{tab:grids}}
\end{table}

\begin{figure}
\centering
\begin{picture}(600,280)
\put(-50,0){\includegraphics[height=4.in]{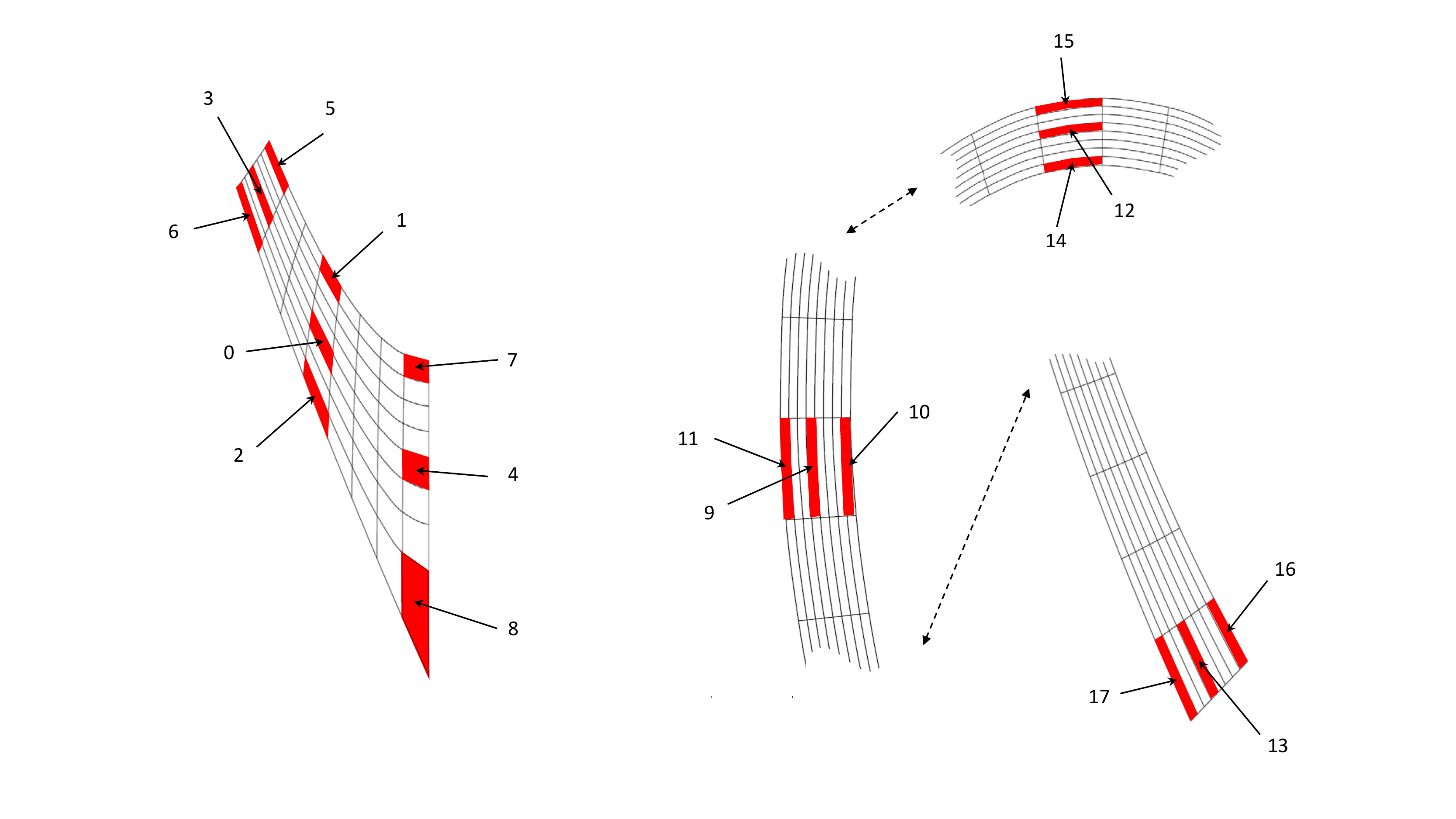}}
\end{picture}
\caption{Test index locations in LCORE (left) and MCORE (right).  Cells 0, 9 and 12 lie
completely within block interiors, cells 1 through 4, 10, 11, 13 and
14 lie on block boundary interiors, and cells 5 though 8, 16 and 17
lie at cell corners.  Of the cells at block boundaries, cells 1, 5, 7,
10, 14 and 16 also lie at physical boundaries.  Cell 8 is adjacent to
the X point. \label{fig:test_patches}} 
\end{figure}

We specify a set $S$ of cells $(i,j)$, indicated in Figure
\ref{fig:test_patches}, relative to the coarsest
($m = 1$) grid in Table \ref{tab:grids}.  We consider only the LCORE and MCORE blocks,
with LCORE representing one of the eight blocks containing the X point
and MCORE representing one of the two blocks that do not.  Further, we consider
only the left side of MCORE due to symmetry.
As can be seen in Figure \ref{fig:test_patches}, cells 0, 9 and 12 lie
completely within block interiors, cells 1 through 4, 10, 11, 13 and
14 lie on block boundary interiors, and cells 5 though 8, 16 and 17
lie at cell corners.  Of the cells at block boundaries, cells 1, 5, 7,
10, 14 and 16 also lie at physical boundaries.  Cell 8 is adjacent to
the X point.  For each
$(i,j)$ in $S$, we compute the velocity space volume-weighted sums
\begin{align}
\tau_{i,j}^{(1)} = \frac{\sum_{k,\ell} \left | r_{i,j,k,\ell}^{(1)} \right |
  V_{i,j,k,\ell}^{(1)}}{\sum_{k,\ell} V_{i,j,k,\ell}^{(1)}},   \label{fig:tau1}
\end{align}
where $r_{i,j,k,\ell}^{(1)}$ is the discrete cell average on phase
space cell $(i,j,k,\ell)$ of the
Vlasov operator applied to the Boltzmann equilibrium solution and
$V_{i,j,k,\ell}^{(1)}$ is the cell volume.  For each $(i,j)$ in $S$,
let $I_{i,j}^{(m)}$ denote the set of configuration
space indices comprised of cells contained in the refinement of cell
$(i,j)$ in grid $m$ for $2 \le m \le 4$.  Letting $V_{i,j,k,\ell}^{(m)}$
again denote the phase
space volumes on the respective grids, define
\begin{align}
\tau_{i,j}^{(m)} = \frac{\sum_{(i',j') \in I_{i,j}^{(m)}} \sum_{k,\ell} \left | r_{i',j',k,\ell}^{(m)} \right |
  V_{i',j',k,\ell}^{(m)}}{\sum_{(i',j') \in I_{i,j}^{(m)}}
  \sum_{k,\ell} V_{i',j',k,\ell}^{(m)}} , ~~~2 \le m \le 4,
\end{align}
which correspond to grid refined evaluations of (\ref{fig:tau1}).

\begin{figure}
\centering
\begin{picture}(600,290)
\put(0,-150){\includegraphics[height=8.in]{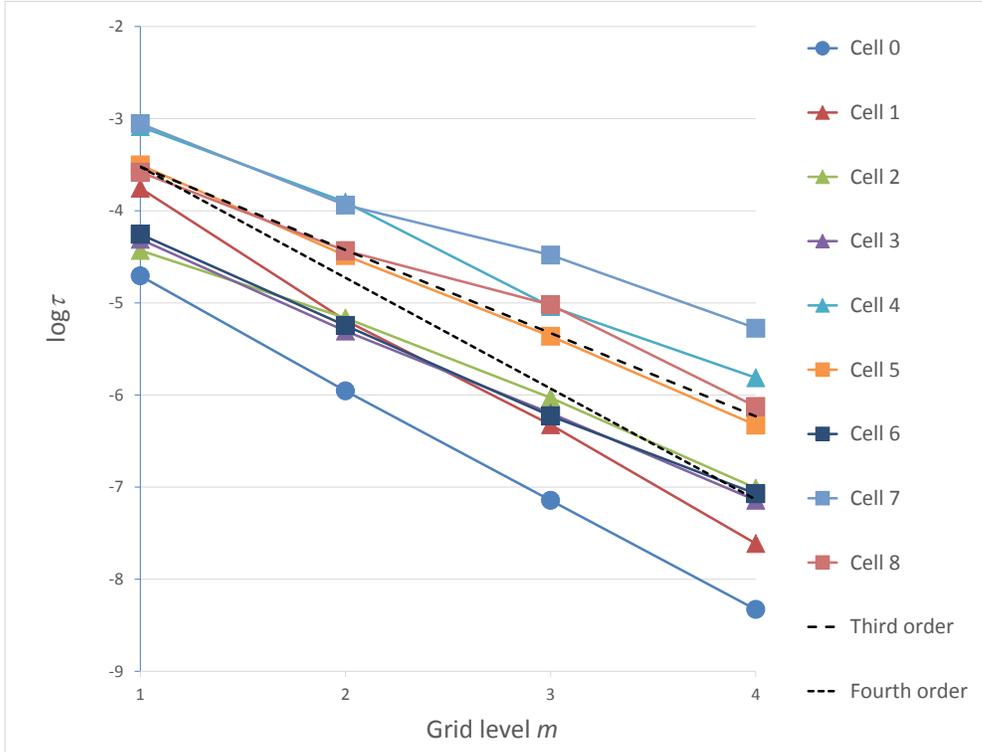}}
\end{picture}
\caption{Truncation error at the LCORE test cells.\label{fig:lcore_error}} 
\end{figure}

\begin{figure}
\centering
\begin{picture}(600,290)
\put(0,-150){\includegraphics[height=8.in]{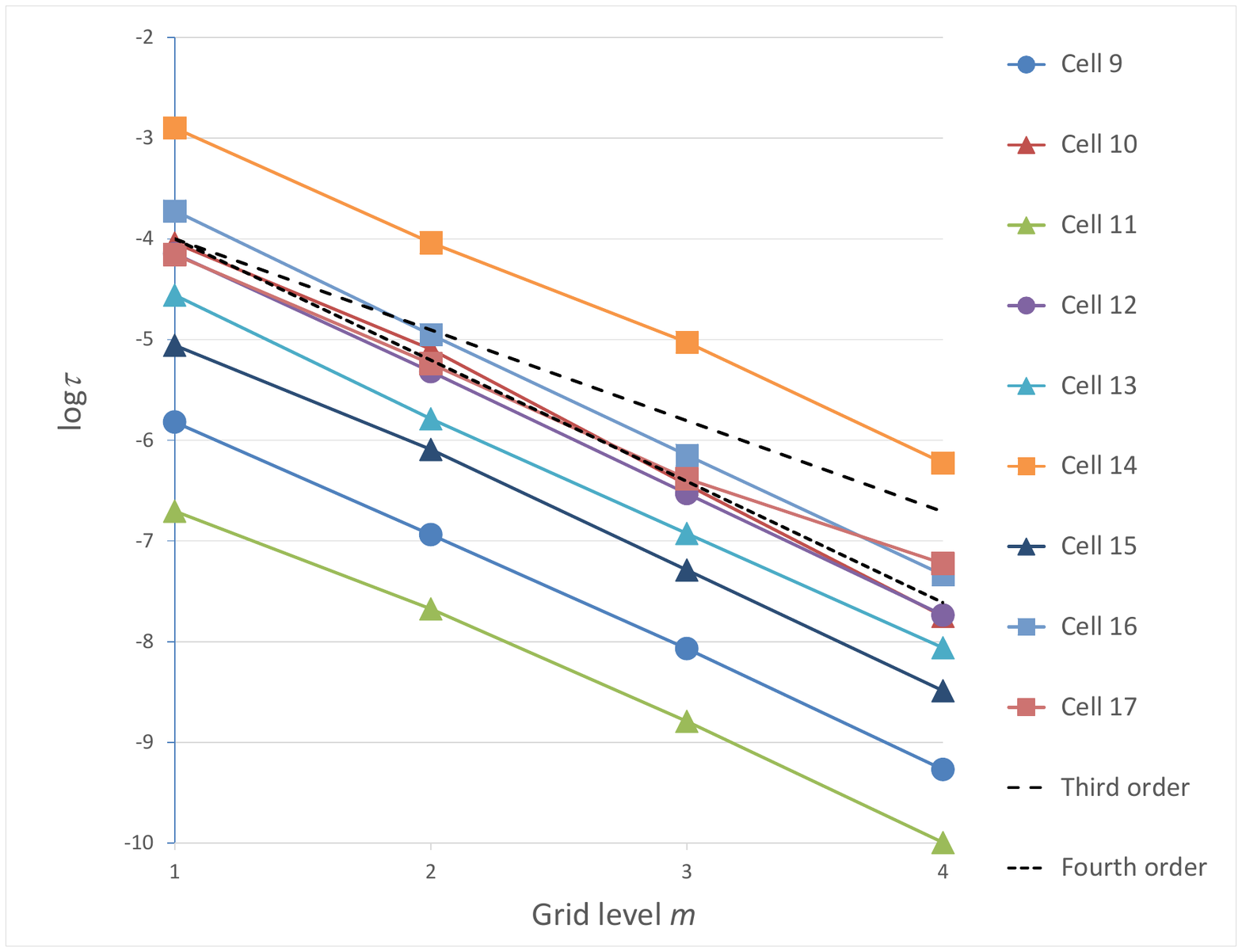}}
\end{picture}
\caption{Truncation error at the MCORE test cells.\label{fig:mcore_error}} 
\end{figure}

Figures \ref{fig:lcore_error} and \ref{fig:mcore_error} are plots of
$\log(\tau_{i,j}^{(m)})$, $1 \le m \le 4$, for each of the 18 test cells
$(i,j)$ enumerated in Figure \ref{fig:test_patches}, along with
reference curves indicating third- and fourth-order convergence.
The curves corresponding to test cells at block interiors, boundaries
and corners are plotted with circular, triangular and square
markers, respectively.  The truncation error at the interior test
cells is seen to converge at fourth-order as
expected.  Since all of the other test cells lie along block
boundaries, we cannot anticipate greater than third-order convergence
in general, yet most of those rates are close to fourth-order as well,
including cell 8, which is adjacent to the X point.

\begin{figure}
\centering
\begin{picture}(600,230)
\put(-40,190){\includegraphics[height=2.5in]{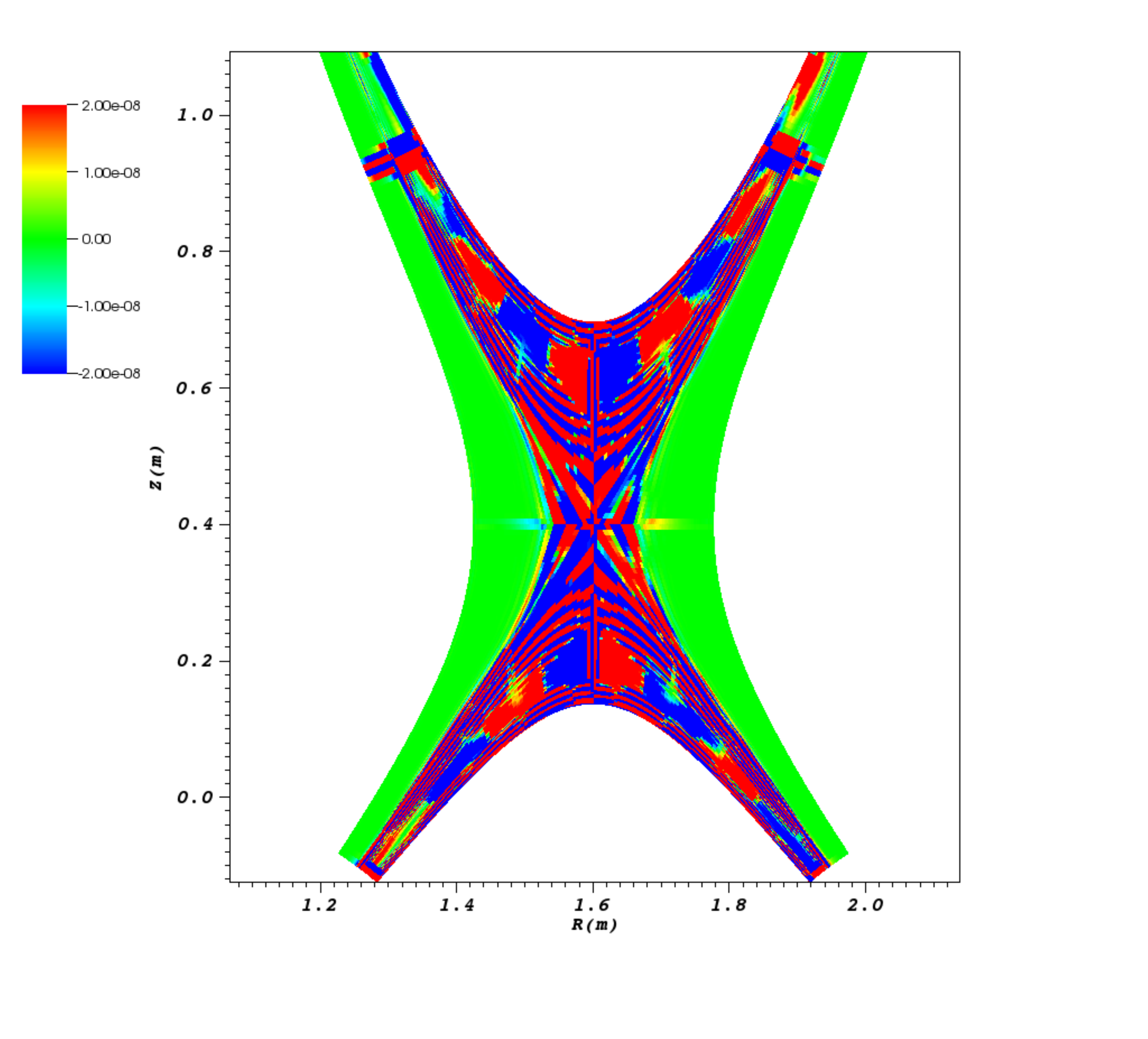}}
\put(130,190){\includegraphics[height=2.5in]{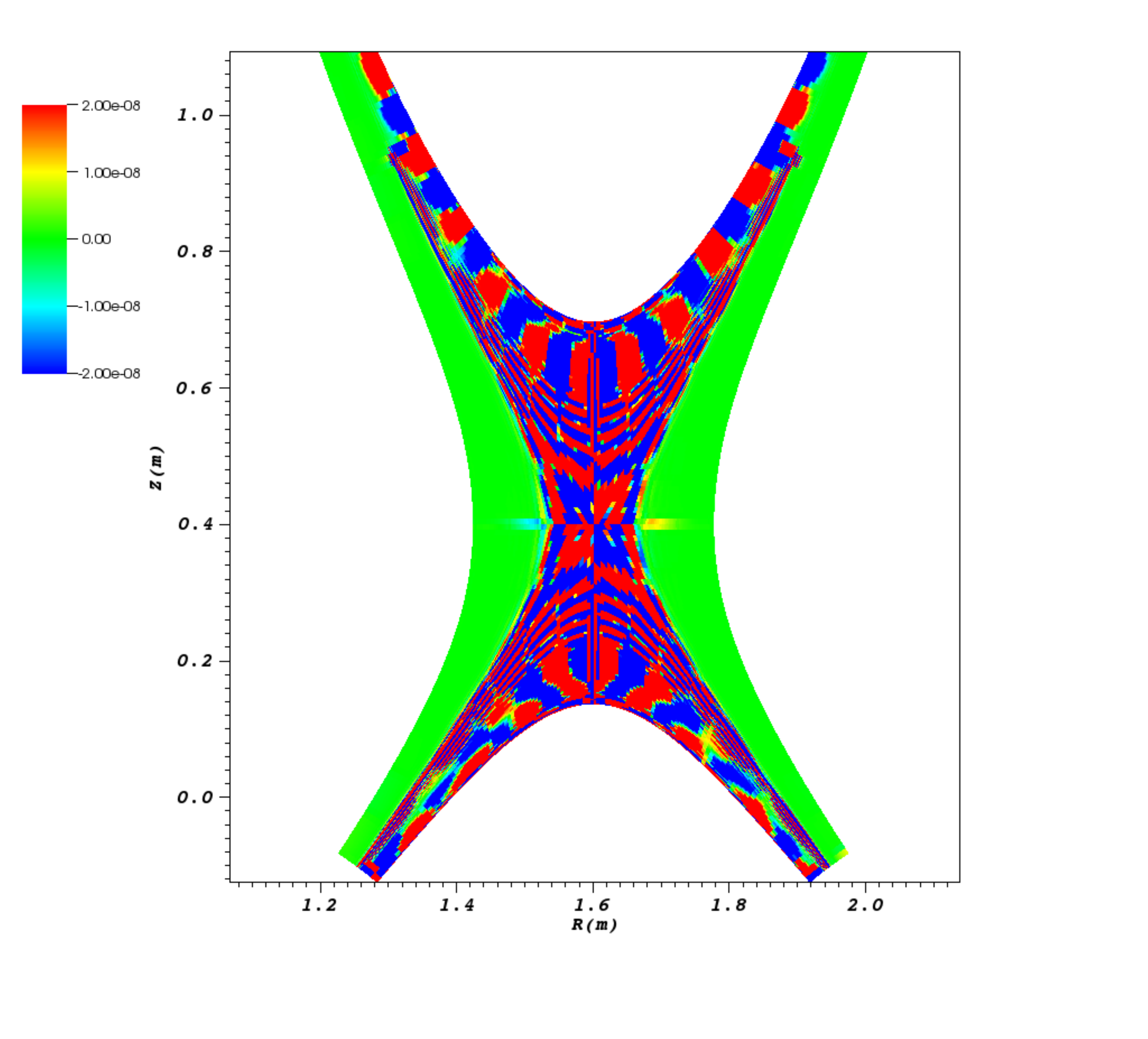}}
\put(300,190){\includegraphics[height=2.5in]{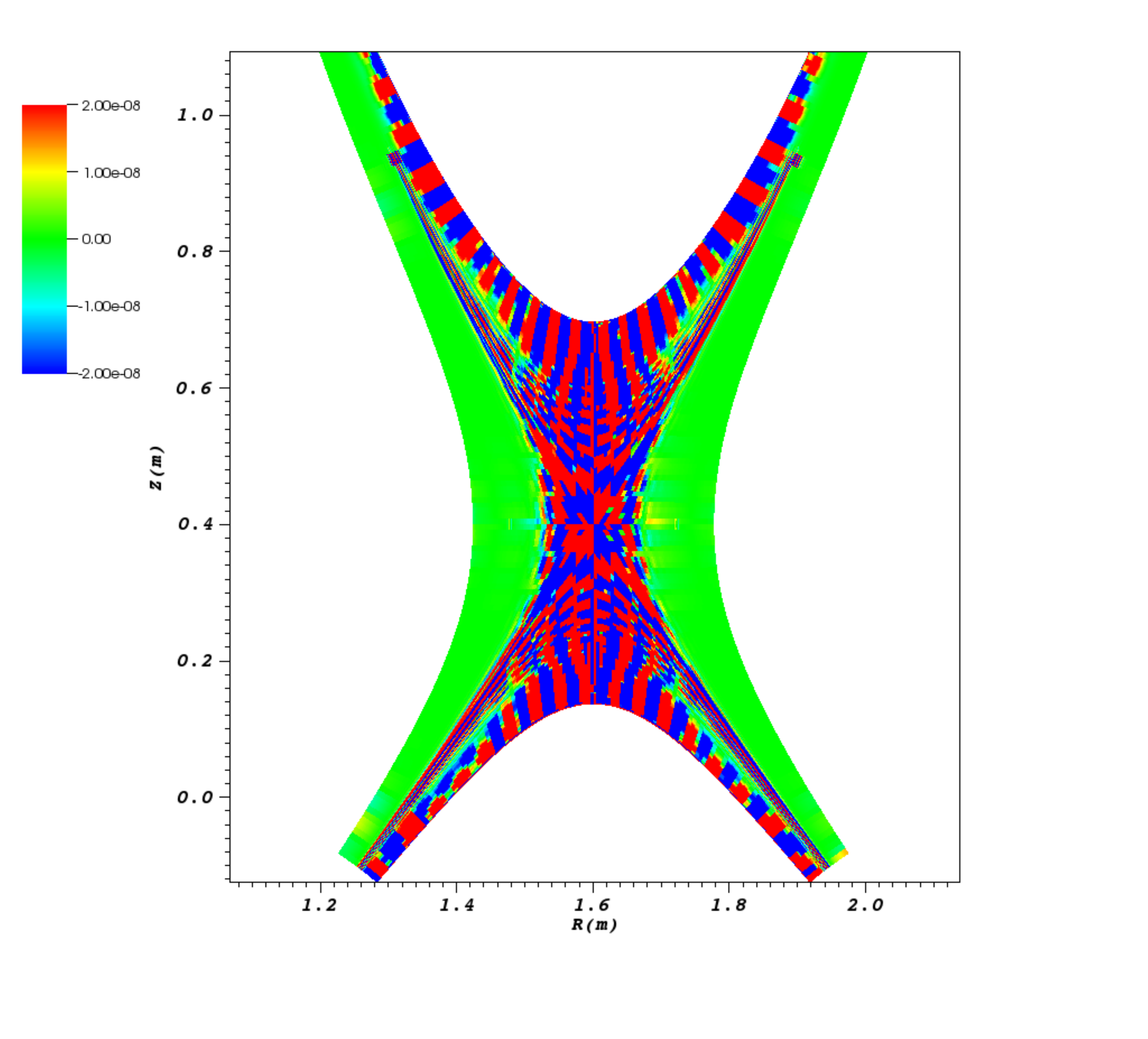}}
\put(30,0){\includegraphics[height=2.5in]{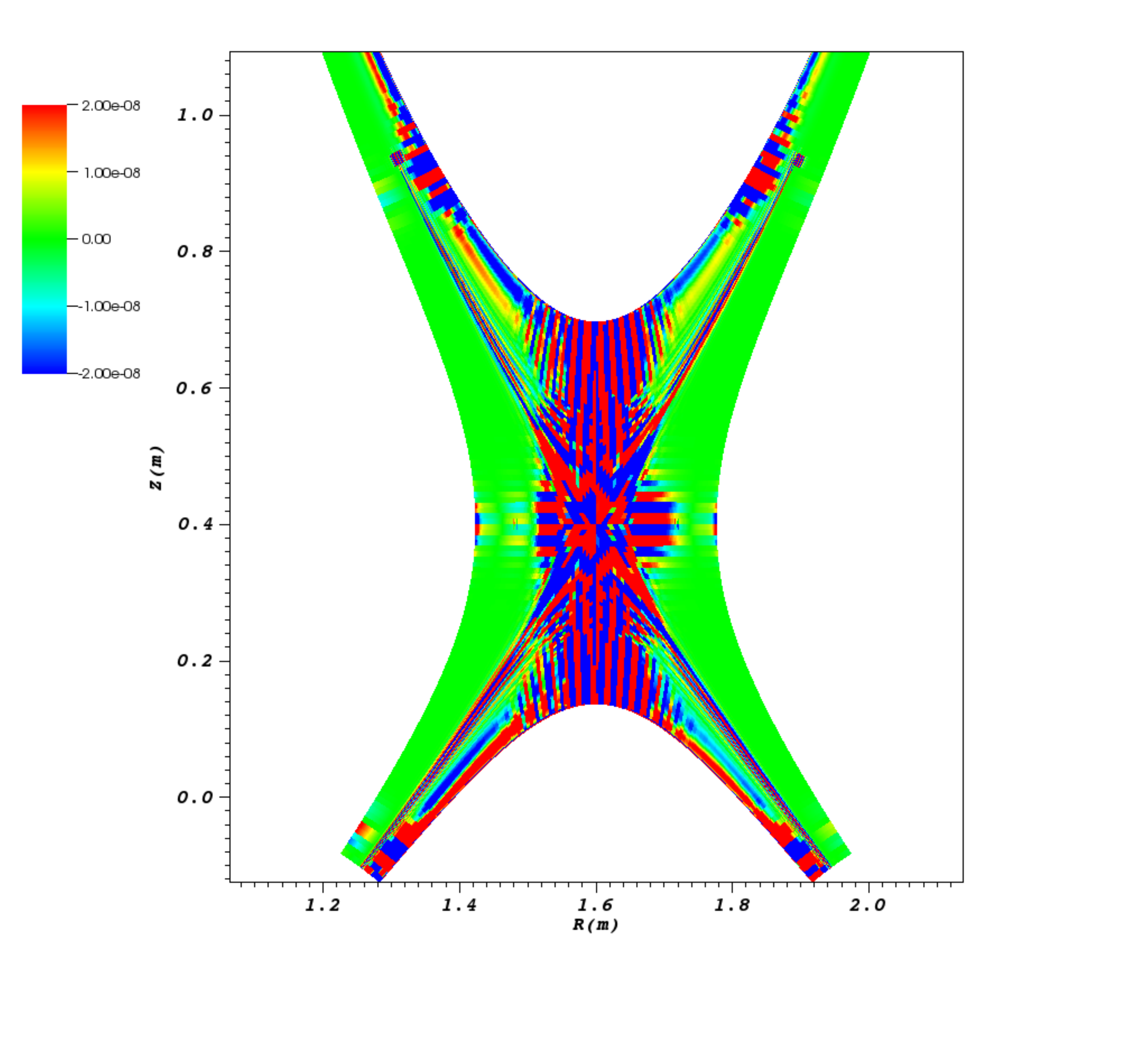}}
\put(220,0){\includegraphics[height=2.5in]{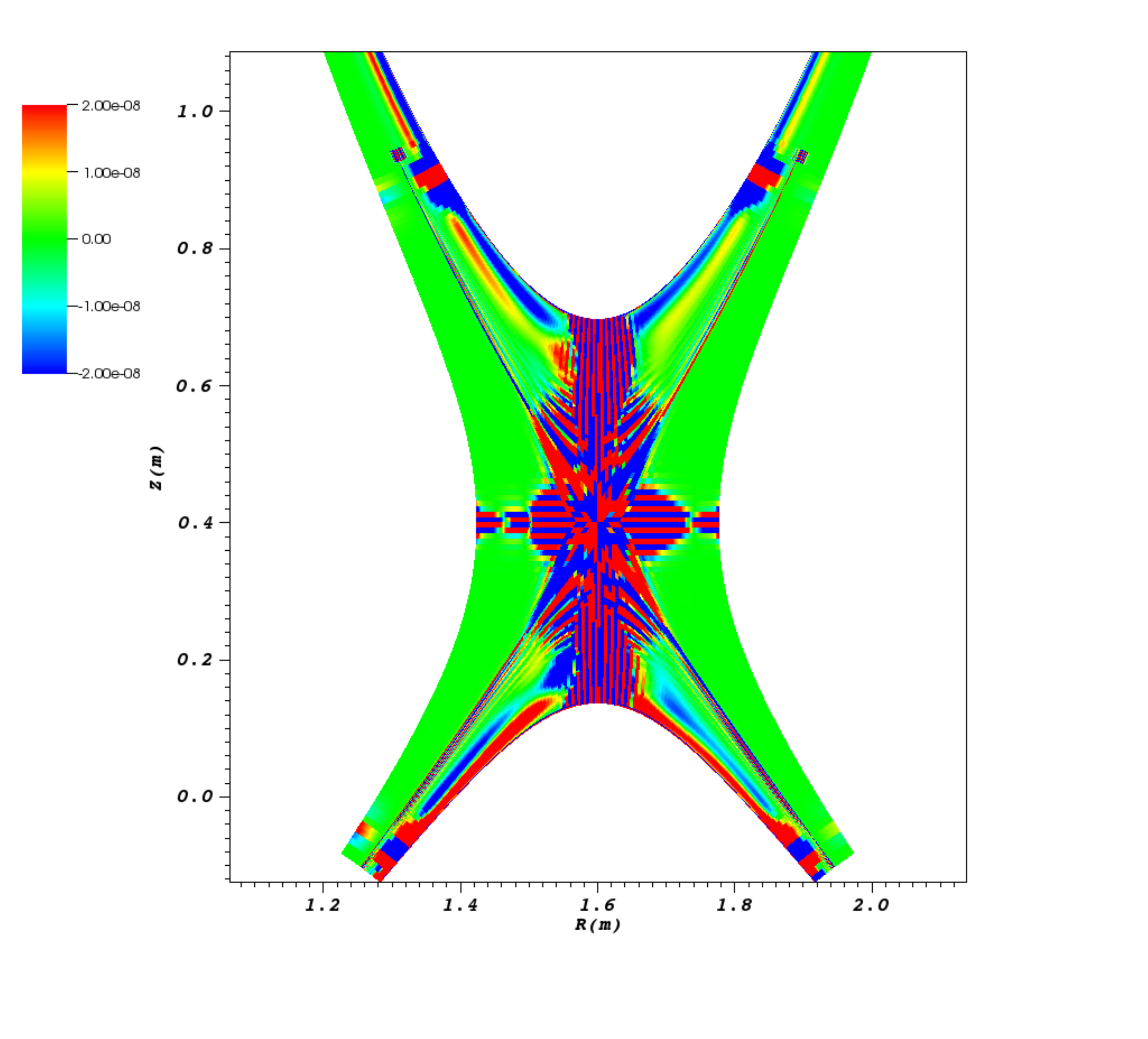}}
\put(10,190){32 poloidal mapping}
\put(185,190){64 poloidal mapping}
\put(350,190){128 poloidal mapping}
\put(80,0){256 poloidal mapping}
\put(270,0){512 poloidal mapping}
\thicklines
\put(335,165){\vector(-1,-1){30}}
\end{picture}
\caption{Truncation error near the X point evaluated at a specified velocity space index as
the poloidal mapping resolution is successively doubled from 32 to 512 cells.  The color maps
  are limited to a particular level ($2 \times
  10^{-8}$) selected to make the error behavior with increasing
  mapping resolution more visible. The arrow in the
  bottom-right plot indicates smooth truncation error associated with the
  density gradient, which emerges after the pollution error
  caused by field misalignment is sufficiently reduced.  Nevertheless,
  larger error due to field dealignment remains near the X point
  relative to the rest of the domain. \label{fig:pollution}} 
\end{figure}

The preceding results were generated using a mapping grid with 24
radial cells in the core blocks and 256 poloidal cells in the union of the core blocks (192
poloidal cells around the MCORE with 32 each in the LCORE and RCORE).
The resolution of the LCSOL, RCSOL, LSOL, RSOL, LPF and RPF blocks was the
same as that of LCORE and MCORE, and the resolution of the MCSOL block
was the same as the MCORE block. The question arises as to how the
mapping grid resolution affects the truncation error in tests like the
one above.  In the approach described in Section \ref{sec:mapping},
the mapping grid nodes are constructed to lie on flux surfaces away
from the X point.  When the poloidal resolution of the computational
grid is finer than that of the mapping grid, the B-spline
interpolation used to evaluate the mapping at points not belonging to
the set of mapping grid nodes is not guaranteed to yield intermediate
points on flux surfaces.  The exact cancellation of the magnetic flux
in the calculation of (\ref{radial_velocity}) will not occur, leading
to the numerical pollution issue described at the end
of Section \ref{sec:divfree}.  To investigate the magnitude of this
effect, we plot in Figure \ref{fig:pollution} the truncation error at
a particular velocity space index and vary the poloidal mapping grid resolution
from 32 to 512 poloidal cells by factors of two.  To make the error
reduction easier to see, the color map in these plots is saturated
at a value of $2 \times 10^{-8}$.  As the poloidal mapping resolution
is increased, the pollution error is sufficiently reduced to observe
the smooth error corresponding to the density gradient in the LCORE
and RCORE blocks, indicated by the arrows in the bottom figure.

\begin{figure}
\centering
\begin{picture}(600,250)
\put(0,-150){\includegraphics[height=8in]{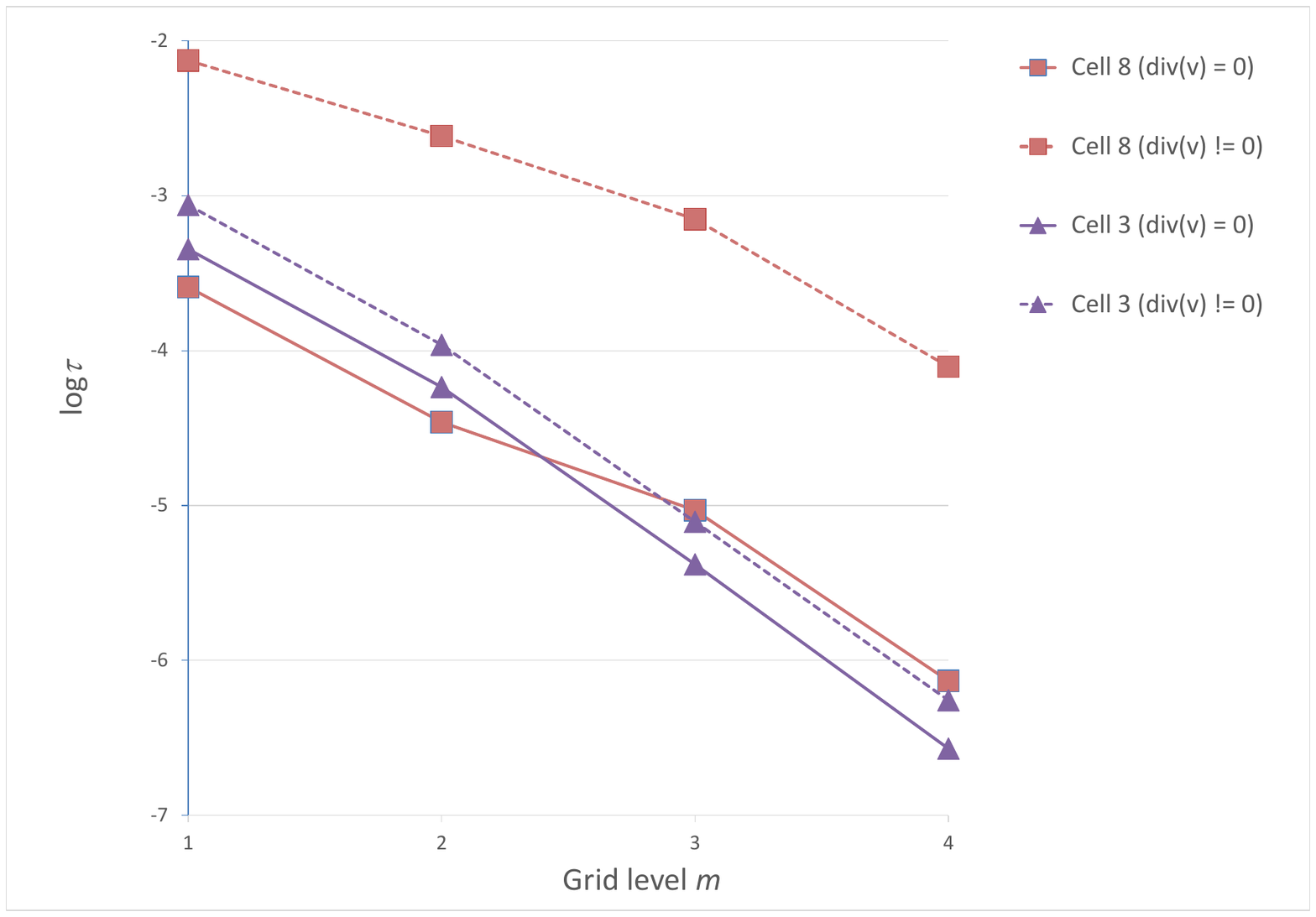}}
\end{picture}
\caption{Comparison of the truncation error at test cells 3 and 8 using the
  divergence-free velocity discretization (solid lines) versus a
  non-divergence-free discretization (dashed lines). \label{fig:vel_compare}} 
\end{figure}

Finally, in Figure \ref{fig:vel_compare}, we plot the truncation error
at test cells 3 and 8 using the divergence-free velocity formulation
described in Section \ref{sec:divfree} (solid lines) versus a
non-divergence-free (albeit fourth-order accurate) velocity
discretization employed in a calculation of fluxes via
(\ref{fddef_fs}) rather than (\ref{fddef_divfree}).  We choose these two cells because they represent
two extremes relative to similar comparisons using the other 16 test
cells.  Whereas the difference between the truncation errors at cell 3
is fairly small, more than an order of magnitude improvement is
observed at cell 8 using the divergence-free formulation.  Error
reductions lying between these two cases are observed for the
remaining test cells.

%% file: summary.tex
\section{Summary \label{summary}}

The predictive simulation of the edge plasma region of a tokamak
fusion reactor via a continuum gyrokinetic model involves multiple
components, including the accurate and efficient advection of plasma
species distribution functions in phase space.  We have described here
our approach for the development and implementation of a conservative,
high-order, MMB, finite-volume, spatial discretization of a full-$f$
gyrokinetic Vlasov model in axisymmetric edge geometries.

An important element of the approach is the discretization of the
gyrokinetic phase space velocity, whose zero divergence property is
preserved discretely to machine precision.  This enables a
conservative formulation of the gyrokinetic system without the
accumulation of truncation error in long-time integration that would
result from a merely asymptotically small velocity divergence.
Furthermore, the discrete velocity contribution of parallel streaming
(the dominant velocity component), $\nabla B$ drifts and curvature
drifts are computed exactly from pointwise evaluations of the magnetic
field data.  Except for a configuration space volume factor (whose
accurate calculation is required in any finite-volume discretization)
in the acceleration terms, no metric factors appear in the velocity
discretization, thereby eliminating the possibility of error
contribution from their discretization, which is an important concern
in mapped grid approaches \cite{Ko06}.  The velocity discretization
respects the asymptotic ordering of the gyrokinetic theory by
eliminating the contribution of the zeroth-order parallel streaming
term on phase space cell faces lying within flux surfaces, except near
the X point where were a flux-aligned grid is precluded, leaving only
the first-order drift terms to be computed without pollution from the
lower-order terms.

We demonstrated the effectiveness of the approach on an
analytically defined edge geometry including both sides of the
magnetic separatrix.  In addition to enabling the direct measurement
of the spatial truncation error resulting from the combination of all
algorithmic components, the use of Boltzmann equilibrium solutions
(\ref{maxwellian_ic}) demonstrates an important advantage of our
approach applied to the edge plasma problem.  As described in Section \ref{sec:intro},
near a tokamak core, a common approximation is to
represent distribution functions as perturbations of a zeroth-order
distribution function, $f = f_0 + \delta \! f$, where $f_0$ is similar
in form to (\ref{maxwellian_ic}).  Since $f_0$ can be computed in the
core, it can be explicitly removed from the model.  Lacking a similar
approach in the edge region, the accurate
calculation of $f_0$ as an integral part of a full-$f$ model
is therefore important to ensure that its discretization error
does not overwhelm the $\delta \! f$ contribution of primary interest.

The spatial discretization described here is the foundation of a
broader effort centered around the development of the COGENT code,
which, as discussed in Section \ref{sec:cogent}, also incorporates
complementary algorithms for the treatment of collision operators,
self-consistent electrostatic fields and
high-order, semi-implicit time integration needed for a complete edge plasma simulation.
To our knowledge, COGENT's ability to solve a
continuum gyrokinetic model in edge geometries spanning both sides of
the magnetic separatrix is unique.
Our current and future work includes the relaxation of the assumption
of toroidal symmetry in the development of a 5D capability for the
simulation of edge turbulence.

%% file: acknowledgments.tex
\section{Acknowledgments \label{acknowledgments}}

We thank Lee Ricketson and Genia Vogman at Lawrence Livermore National
Laboratory for their helpful comments and suggestions.
This material is based on work supported by the U.S. Department of
Energy, Office of Science, Office of Advanced Scientific Computing
Research, Applied Mathematics Program under Contract DE-AC52-07NA27344
at Lawrence Livermore National Laboratory and Contract
DE-AC02-05CH11231 at Lawrence Berkeley National Laboratory.

%% file: appendix.tex
\section{Appendix \label{sec:appendix}}

\subsection{Normalizations \label{normalizations}}
Equations (\ref{conservativenormalized})-(\ref{gkvelocity_vars}) and
(\ref{gkpoisson}) are obtained from the gyrokinetic Vlasov-Poisson system derived
in \cite{Ha96} using a normalization relative to a reference material and fields described by
the primitive parameters specified in Table~\ref{primitiverefs} and the derived quantities in Table~\ref{derivedrefs}.
\begin{table}
\center
\begin{tabular}{|c|l|}
\hline
$\widetilde{n}$ & number density \\
$\widetilde{T}$ & temperature \\ 
$\widetilde{L}$ & length \\
$\widetilde{m}$ & mass \\
$\widetilde{B}$ & magnetic field \\
\hline
\end{tabular}
\caption{Primitive reference parameters.} \label{primitiverefs}
\end{table}
\begin{table}[h]
\center
\begin{tabular}{|l|l|}
\hline
$\widetilde{v} = \left ( \widetilde{T} / \widetilde{m} \right )^{1/2}$ & thermal speed \\
$\widetilde{\tau} = \widetilde{L} / \widetilde{v}$ & transit time \\
$\widetilde{\mu} = \widetilde{T}/(2\widetilde{B})$ & magnetic moment \\
$\widetilde{f} = \widetilde{n} / (\pi \widetilde{v}^3 )$ &
distribution function \\
$\widetilde{\Phi} = \widetilde{T} / e $ & potential\\
$\widetilde{\Omega} = e \widetilde{B} / \widetilde{m}$ &
gyrofrequency \\
$\widetilde{\lambda}_D = \left ( \epsilon_0 \widetilde{T} /
(\widetilde{n} e^2 ) \right )^{1/2} $ & Debye length \\
\hline
\end{tabular}
\caption{Derived reference parameters.} \label{derivedrefs}
\end{table}
The normalized variables used in the gyrokinetic Vlasov-Poisson model are
displayed in Table \ref{table:normalizedvars}.
\begin{table}
\center
\begin{tabular}{|l|l|}
\hline
$\widehat{t} = t / \widetilde{\tau} $ & time \\
$\widehat{v}_{\parallel} = v_{\parallel} /
\widetilde{v}$ & parallel velocity \\
$\widehat{n}_{\alpha} = n_{\alpha} / \widetilde{n}$ & number density \\
%$\widehat{v}_{j} = v_{j} / \widetilde{v}$ & velocity \\
%$\widehat{v}_{\alpha,\perp} = v_{\alpha,\perp} /
%\widetilde{v}$ & perpendicular velocity \\
$\widehat{m}_{\alpha} = m_{\alpha} / \widetilde{m}$  & mass \\
$\widehat{f}_{\alpha} = f_{\alpha} / \widetilde{f}$ & distribution
function \\
$\widehat{T}_{\alpha} = T_{\alpha} / \widetilde{T}$ & temperature
\\
$\widehat{B} = B / \widetilde{B} $ & magnetic field \\
$\widehat{\Phi} = \Phi / \widetilde{\Phi}$  & potential \\
$\widehat{\mu} = \mu / \widetilde{\mu} $ & magnetic moment \\
\hline
\end{tabular}
\caption{Normalized gyrokinetic Vlasov-Poisson variables.} \label{table:normalizedvars}
\end{table}
In terms of the reference scales, two dimensionless numbers, defined
in Table~\ref{table:dimlessnums}, appear in the normalized gyrokinetic Vlasov-Poisson system.
\begin{table}[t]
\center
\begin{tabular}{|l|l|}
\hline
$\rho_L = \widetilde{v} / (\widetilde{\Omega} \widetilde{L}) $ &
Larmor number: ratio of gyroradius to scale length\\
$\lambda_D = \widetilde{\lambda}_D/\widetilde{L}$ & Debye number: ratio of Debye
length to scale length \\
\hline
\end{tabular}
\caption{Dimensionless gyrokinetic Vlasov-Poisson parameters.} \label{table:dimlessnums}
\end{table}

For the magnetic flux in edge geometries, we employ the normalization
$\psi_N = (\psi - \psi_A) / (\psi_S - \psi_A)$,
where $\psi_A$ and $\psi_S$ are the on-axis and separatrix flux
values, respectively.

\subsection{Gyrokinetic velocity divergence \label{sec:veldiv}}

Define $\mbu = (\mbu_\mbR,u_{\vpll}) = (\Bpll^*
\dot{\mbR},\Bpll^* \dot{\vpll})$.  Letting $\mbA$ be such that $\mbB =
\nabla_{\mbR} \times \mbA$ and defining $p$ as in (\ref{pdef}),
we have from (\ref{gkvelocity})-(\ref{gkvelocity_vars}) that
\begin{subequations}
\begin{align}
\mbu_{\mbR} &= \nabla_{\mbR} \times \left \{ \vpll \left ( \mbA + \frac{
  m \vpll\rho_L}{Z} \mbb \right )
  \right \} + \mbb \times \nabla_{\mbR} p, \\
u_{\vpll} &= - \nabla_{\mbR} \times \left ( \frac{Z}{m\rho_L} \mbA + 
  \vpll \mbb \right ) \cdot \nabla_{\mbR} p.
\end{align}
\end{subequations}
Therefore,
\begin{align}
\nabla_{\mbR} \cdot \mbu_{\mbR} &= \nabla_{\mbR} \cdot \left ( \mbb
\times \nabla_{\mbR} p \right ) \label{divuR} \\
\nonumber &= \nabla_{\mbR} p \cdot \left ( \nabla_{\mbR} \times \mbb \right ) -
\mbb \cdot \left ( \nabla_{\mbR}
\times \nabla_{\mbR} p \right ) \\
\nonumber &= \nabla_{\mbR} p \cdot \left ( \nabla_{\mbR} \times \mbb \right ),
\end{align}
and
\begin{align}
\frac{\partial u_{\vpll}}{\partial \vpll}
= - \left ( \nabla_{\mbR} \times \mbb \right ) \cdot \nabla_{\mbR} p,
\end{align}
from which (\ref{veldiv}) follows.

\subsection{Face integrals \label{section:face_integrals}}

In direction 1, the face integral (\ref{normalvelint}) for $\alpha = 0$ or $1$ is
evaluated as:
\begin{align}
  \int \limits_{\mbX(V^\alpha_1)} & \bsigma  = \\
& \int \limits_{V^\alpha_1} \left ( \widetilde{u}_3 \det
  \begin{array}{|ccc|}
    \frac{\partial \mbX_0}{\partial \xi_0} & \frac{\partial \mbX_0}{\partial \xi_2}  & \frac{\partial \mbX_0}{\partial \xi_3}  \\
\nonumber     \frac{\partial \mbX_1}{\partial \xi_0} & \frac{\partial \mbX_1}{\partial \xi_2}  & \frac{\partial \mbX_1}{\partial \xi_3}  \\
\nonumber     \frac{\partial \mbX_2}{\partial \xi_0} & \frac{\partial \mbX_2}{\partial \xi_2}  & \frac{\partial \mbX_2}{\partial \xi_3}
\end{array} ~\right )_{\xi_1 = \xi_1^\alpha} \hspace{-2em}
d\xi_0 d\xi_2 d\xi_3 
 - \int \limits_{V^\alpha_1} \left ( \widetilde{u}_2 \det
  \begin{array}{|ccc|}
    \frac{\partial \mbX_0}{\partial \xi_0} & \frac{\partial \mbX_0}{\partial \xi_2}  & \frac{\partial \mbX_0}{\partial \xi_3}  \\
    \frac{\partial \mbX_1}{\partial \xi_0} & \frac{\partial \mbX_1}{\partial \xi_2}  & \frac{\partial \mbX_1}{\partial \xi_3}  \\
    \frac{\partial \mbX_3}{\partial \xi_0} & \frac{\partial \mbX_3}{\partial \xi_2}  & \frac{\partial \mbX_3}{\partial \xi_3}
\end{array} ~\right )_{\xi_1 = \xi_1^\alpha} \hspace{-2em}
d\xi_0 d\xi_2 d\xi_3 \\
\nonumber + & \int \limits_{V^\alpha_1} \left ( \widetilde{u}_1 \det
  \begin{array}{|ccc|}
    \frac{\partial \mbX_0}{\partial \xi_0} & \frac{\partial \mbX_0}{\partial \xi_2}  & \frac{\partial \mbX_0}{\partial \xi_3}  \\
\nonumber     \frac{\partial \mbX_2}{\partial \xi_0} & \frac{\partial \mbX_2}{\partial \xi_2}  & \frac{\partial \mbX_2}{\partial \xi_3}  \\
\nonumber     \frac{\partial \mbX_3}{\partial \xi_0} & \frac{\partial \mbX_3}{\partial \xi_2}  & \frac{\partial \mbX_3}{\partial \xi_3}
\end{array} ~\right )_{\xi_1 = \xi_1^\alpha} \hspace{-2em}
d\xi_0 d\xi_2 d\xi_3 
 - \int \limits_{V^\alpha_1} \left ( \widetilde{u}_0 \det
  \begin{array}{|ccc|}
    \frac{\partial \mbX_1}{\partial \xi_0} & \frac{\partial \mbX_1}{\partial \xi_2}  & \frac{\partial \mbX_1}{\partial \xi_3}  \\
\nonumber     \frac{\partial \mbX_2}{\partial \xi_0} & \frac{\partial \mbX_2}{\partial \xi_2}  & \frac{\partial \mbX_2}{\partial \xi_3}  \\
\nonumber     \frac{\partial \mbX_3}{\partial \xi_0} & \frac{\partial \mbX_3}{\partial \xi_2}  & \frac{\partial \mbX_3}{\partial \xi_3}
\end{array} ~\right )_{\xi_1 = \xi_1^\alpha} \hspace{-2em}
d\xi_0 d\xi_2 d\xi_3 \\
\nonumber = & \int \limits_{V^\alpha_1} \left ( \widetilde{u}_3 \det
  \begin{array}{|ccc|}
    1 & 0 & 0  \\
    0 & \frac{\partial \mbX_1}{\partial \xi_2}  & \frac{\partial \mbX_1}{\partial \xi_3}  \\
    0 & \frac{\partial \mbX_2}{\partial \xi_2}  & \frac{\partial \mbX_2}{\partial \xi_3}
\end{array} ~\right )_{\xi_1 = \xi_1^\alpha} \hspace{-2em}
d\xi_0 d\xi_2 d\xi_3
- \int \limits_{V^\alpha_1} \left ( \widetilde{u}_2 \det
  \begin{array}{|ccc|}
    1 & 0 & 0 \\
    0 & \frac{\partial \mbX_1}{\partial \xi_2}  & \frac{\partial \mbX_1}{\partial \xi_3}  \\
    0 & \frac{\partial \mbX_3}{\partial \xi_2}  & \frac{\partial \mbX_3}{\partial \xi_3}
\end{array} ~\right )_{\xi_1 = \xi_1^\alpha} \hspace{-2em}
d\xi_0 d\xi_2 d\xi_3 \\
\nonumber + & \int \limits_{V^\alpha_1} \left ( \widetilde{u}_1 \det
  \begin{array}{|ccc|}
    1 & 0 & 0 \\
    0 & \frac{\partial \mbX_2}{\partial \xi_2}  & \frac{\partial \mbX_2}{\partial \xi_3}  \\
    0 & \frac{\partial \mbX_3}{\partial \xi_2}  & \frac{\partial \mbX_3}{\partial \xi_3}
\end{array} ~\right )_{\xi_1 = \xi_1^\alpha} \hspace{-2em}
d\xi_0 d\xi_2 d\xi_3
 - \int \limits_{V^\alpha_1} \left ( \widetilde{u}_0 \det
  \begin{array}{|ccc|}
    0 & \frac{\partial \mbX_1}{\partial \xi_2}  & \frac{\partial \mbX_1}{\partial \xi_3}  \\
    0 & \frac{\partial \mbX_2}{\partial \xi_2}  & \frac{\partial \mbX_2}{\partial \xi_3}  \\
    0 & \frac{\partial \mbX_3}{\partial \xi_2}  & \frac{\partial \mbX_3}{\partial \xi_3}
\end{array} ~\right )_{\xi_1 = \xi_1^\alpha} \hspace{-2em}
d\xi_0 d\xi_2 d\xi_3 \\
\nonumber = & \int \limits_{V^\alpha_1} \left ( \widetilde{u}_3 N_{1,3}^T + \widetilde{u}_2 N_{1,2}^T + \widetilde{u}_1 N_{1,1}^T \right )_{\xi_1 = \xi_1^\alpha}
d\xi_0 d\xi_2 d\xi_3 \\
\nonumber = & \int \limits_{V^\alpha_1} \left ( \mbN^T \widetilde{\mbu} \right )_1 d\mbV_{\bxi}.
\end{align}
The integrals (\ref{normalvelint}) in the other directions are obtained similarly.

\subsection{Identification of the skew-symmetric tensor used in the
  gyrokinetic velocity discretization \label{sec:veldecomp}}

\noindent
To find a skew-symmetric, second-order tensor satisfying
(\ref{divtensor})-(\ref{zetadef}), we note that 4-vectors of the form $(0,
\nabla_{\boldsymbol{R}} \times \mbv)$ with $\mbv = (v_1,v_2,v_3)$ are trivially
expressible as a skew-symmetric tensor divergence:
\begin{align}
\left ( 0, \nabla_{\boldsymbol{R}} \times \mbv \right )_j = \sum_{j'=0}^3
\frac{\partial \eta_{j,j'}}{\partial x_{j'}}, ~~ 0 \le j \le 3,
\end{align}
where
\begin{align}
\eta = \left (
\begin{array}{cccc}
0 & 0 & 0 & 0 \\
0 & 0 & v_3 & -v_2 \\
0 & -v_3 & 0 & v_1 \\
0 & v_2 & -v_1 & 0
\end{array}
\right ).
\end{align}
We therefore separate the
terms of (\ref{gkvelocity}) involving $\mbB
= \nabla_{\boldsymbol{R}} \times \mbA$ and $\nabla_{\boldsymbol{R}}
\times \mbb$, resulting in
a decomposition $\widetilde{\mbu} =
\widetilde{\mbu}_1 + \widetilde{\mbu}_2$, where
\begin{subequations}
\begin{align}
\widetilde{\mathbf{u}}_1 &= \left ( 0, \nabla_{\boldsymbol{R}} \times \left [ \vpll \left ( \mbA + \frac{
  m \vpll\rho_L}{Z} \mbb  \right ) \right ] \right ),\\
\widetilde{\mathbf{u}}_2 &= \left ( -v_{\parallel} (\nabla_{\boldsymbol{R}} \times \mbb ) \cdot
\nabla_{\boldsymbol{R}} p , \mbb \times \nabla_{\boldsymbol{R}} p \right ).
\end{align}
\end{subequations}
The entries $\omega_{i,j}, 1 \leq i < j \leq 3,$ given by
(\ref{omega12})-(\ref{omega23}) are then obtained from $\widetilde{\mathbf{u}}_1$.
We obtain the contribution from $\widetilde{\mathbf{u}}_2$
by setting $\omega_{0,j} = (-1)^{j+1} v_\parallel ( \mbb \times
  \nabla_{\boldsymbol{R}} p )_j$, $1 \le j \le 3$, since, using (\ref{zetadef}) and (\ref{divuR}),
\begin{subequations}
\begin{align}
\nabla_{\boldsymbol{R}} \cdot (\zeta_{0,1},\zeta_{0,2},\zeta_{0,3} ) &=
\nabla_{\boldsymbol{R}} \cdot (-\omega_{0,1},\omega_{0,2},-\omega_{0,3}
) = -v_\parallel (\nabla_{\boldsymbol{R}} \times \mbb ) \cdot
\nabla_{\boldsymbol{R}} p , \\
  \frac{\partial}{\partial v_\parallel}
  (\zeta_{0,1},\zeta_{0,2},\zeta_{0,3} )
 &=  \frac{\partial}{\partial v_\parallel}
  (-\omega_{0,1},\omega_{0,2},-\omega_{0,3} ) 
= \mbb \times \nabla_{\boldsymbol{R}} p .
\end{align}
\end{subequations}

\subsection{An inequality used in the construction of mapping grids \label{sec:inequality}}

In the derivation of (\ref{mapcoefs}), assuming that $b \ne 0$ and $b^2 > 4ac$, we have
\begin{align}
\left [ b^2 - 2c(a-c) \right ]^2 - \left [ 2c \sqrt{b^2
 + (a-c)^2} \right ]^2 = b^2 ( b^2 - 4ac) > 0,
\end{align}
so
\begin{align}
2 |c| \sqrt{ b^2 + (a-c)^2} < \left | b^2 - 2c (a-c) \right |. \label{ineq2}
\end{align}
However,
\begin{align}
b^2 - 2c (a-c) > 0,  \label{ineq1}
\end{align}
since, if $ac \le 0$, (\ref{ineq1}) is verified trivially; otherwise
\begin{align}
b^2 - 2c(a-c) = b^2 - 2ac +2c^2 > b^2 -4ac + 2c^2 > 0.
\end{align}
Therefore, using (\ref{ineq2}),
\begin{align}
-2 c \sqrt{ b^2 + (a-c)^2} \le 2 |c| \sqrt{ b^2 + (a-c)^2} < b^2 - 2c (a-c),
\end{align}
yielding
\begin{align}
b^2 + 2c(-a + \sqrt{b^2 + (a-c)^2} + c) > 0.
\end{align}

\subsection{Boltzmann equilibrium \label{boltzmann_eq}}

With the definitions (\ref{maxwellian_ic})-(\ref{potential_ic}) and
\begin{align}
w = \pi^{-1/2} (2T/m)^{-3/2} \exp \left ( - \frac{m \vpll^2 + \mu B(\psi,\theta)}{2T} \right ),
\end{align}
we have
\begin{subequations}
\begin{align}
  \bnabla f &= w \left ( - \frac{n\mu}{2T} \bnabla B + \bnabla n \right ), \\
  \partial_{\vpll} f &= - w\frac{m\vpll n}{T}, \\
  \bnabla \phi &= - \frac{T}{Zn} \bnabla n.
\end{align}
\end{subequations}
Since $\mbB \cdot \bnabla n = \mbB \cdot \bnabla \phi = 0$, we have
\begin{subequations}
\begin{align}
  \mbB^* \cdot \bnabla f &= w \left [ - \frac{n\mu}{2T} \mbB \cdot \bnabla B + \rho_L \frac{m \vpll}{Z} (\bnabla \times \mbb) \cdot \left ( - \frac{n\mu}{2T} \bnabla B + \bnabla n \right ) \right ], \\
  \mbB^* \cdot \mbG &= \frac{\mu}{2} \mbB \cdot \bnabla B - \rho_L \frac{m \vpll}{Z} \left ( \bnabla \times \mbb \right ) \cdot \left ( \frac{T}{n} \bnabla n - \frac{\mu}{2} \bnabla B \right ),
\end{align}
\end{subequations}
and hence
\begin{align}
\mbB^* \cdot \left ( \vpll \bnabla f + w\frac{\vpll n}{T} \mbG \right ) = 0.
\end{align}
Also,
\begin{align}
\mbG \times \bnabla f = w \left [ \frac{\mu}{2} \bnabla n \times \bnabla B - \frac{T}{n} \bnabla n \times \bnabla n - \frac{n \mu^2}{4T}\bnabla B \times \bnabla B + \frac{\mu}{2} \bnabla B \times \bnabla n \right ] = 0.
\end{align}
Therefore,
\begin{align}
  \Bpll^* \left ( \dot{\mbR} \cdot \bnabla f + \dot{\vpll} \partial_{\vpll} f \right ) &= \mbB^* \cdot \left ( \vpll \bnabla f + w\frac{\vpll n}{T} \mbG \right ) + \frac{\rho_L}{2} \left ( \mbb \times \mbG \right ) \cdot \bnabla f \\
\nonumber  &= \frac{\rho_L}{2} \left ( \mbG \times \bnabla f \right ) \cdot \mbb \\
\nonumber  &= 0.
\end{align}
Since the gyrokinetic velocity is divergence free (Appendix \ref{sec:veldiv}),
\begin{equation}
  \bnabla \cdot  \left( \dot{\mbR} \Bpll^* f \right) 
  + \frac{\partial}{\partial v_{\parallel}} \left(
    \dot{\vpll} \Bpll^* f \right) = 0.
\end{equation}